\documentclass[useAMS,usenatbib,aps,floatdix,nofootinbib]{mn2e}

\usepackage{graphics}
\usepackage{graphicx}

\usepackage{amssymb,amsmath}

\usepackage{amsmath}
\usepackage{psfrag}
\usepackage{graphicx}

\usepackage{color}
\usepackage{ulem}

\newcommand{\f}{\frac}

\newcommand{\be}{\begin{equation}}      
\newcommand{\ee}{\end{equation}}      
      
\newcommand{\bef}{\begin{figure*}}      
\newcommand{\eef}{\end{figure*}}      
\newcommand{\bea}{\begin{eqnarray}}    
\newcommand{\eea}{\end{eqnarray}}      
  
\def\spose#1{\hbox to 0pt{#1\hss}}
\def\ltapprox{\mathrel{\spose{\lower 3pt\hbox{$\mathchar"218$}}
\raise 2.0pt\hbox{$\mathchar"13C$}}}
\def\gtapprox{\mathrel{\spose{\lower 3pt\hbox{$\mathchar"218$}}
\raise 2.0pt\hbox{$\mathchar"13E$}}}
\def\inapprox{\mathrel{\spose{\lower 3pt\hbox{$\mathchar"218$}}
\raise 2.0pt\hbox{$\mathchar"232$}}}

\def\bse{\begin{subequations}}
\def\ese{\end{subequations}}

\def\lsim{\raise 0.4ex\hbox{$<$}\kern -0.8em\lower 0.62ex\hbox{$\sim$}} 
\def\gsim{\raise 0.4ex\hbox{$>$}\kern -0.7em\lower 0.62ex\hbox{$\sim$}}

\def\f0N{f_0^{(N)}}
\def\bec{\begin{center}}
\def\eec{\end{center}}

\usepackage{xspace}
\newcommand{\Abacus}{\textsc{Abacus}\xspace}

\title[Quantifying resolution of cosmological simulations]
{Quantifying resolution in cosmological $N$-body simulations using self-similarity}
\author[M. Joyce et al.]
{Michael Joyce${^{1}}$\thanks{Corresponding author email: joyce@lpnhe.in2p3.fr}, Lehman Garrison${^{2,3}}$ and Daniel Eisenstein${^{2}}$ \\
$^1$Laboratoire de Physique Nucl\'eaire et de Hautes \'Energies, UPMC IN2P3 CNRS UMR 7585, \\
\,\,\,Sorbonne Universit\'e, 4, place Jussieu, 75252 Paris Cedex 05, France \\\
$^2$Harvard \& Smithsonian Center for Astrophysics, 60 Garden St, Cambridge, MA 02138 \\
$^3$Center for Computational Astrophysics, Flatiron Institute, 162 Fifth Ave., New York, NY 10010}
\begin{document}

\date{\today}

\maketitle

\begin{abstract}
We demonstrate that testing for self-similarity in scale-free simulations provides an excellent tool to quantify the resolution at small scales of cosmological N-body simulations.
Analysing two-point correlation functions measured in simulations using \Abacus, we show how observed deviations from self-similarity reveal the range of time and distance scales in which convergence is obtained.
While the well-converged scales show accuracy below 1\%, our results show that, with a small force softening length, the spatial resolution is essentially determined  by the mass resolution.
At later times the lower cut-off scale on convergence evolves 
in comoving units as $a^{-1/2}$ ($a$ being the scale factor), consistent with a hypothesis that it is set by two-body collisionality. A corollary
of our results is that N-body simulations, particularly at high
red-shift, contain a significant spatial range in which clustering
appears converged with respect to the time-stepping and force softening but has not actually converged to the physical continuum result.
The method developed can be applied {to determine the resolution of any clustering statistic} and extended to infer resolution limits for non-scale-free simulations.

\end{abstract}


\section{Introduction} 

As the quantity and quality of observational data will increase dramatically over the coming years, the need for theoretical model predictions 
that are sufficiently precise in order to fully exploit these data has become an urgent issue. In practice the exclusive tool to calculate predictions for the non-linear regime of cosmological structure formation remains numerical simulation based on the $N$-body method, in which the evolution of the phase-space density of dark matter is approximated by following that of a finite particle sampling. This method imposes evidently an intrinsic limit on the spatial (or mass) resolution of the results obtained. However, because of the complexity of the non-linear gravitational dynamics, the quantification of this resolution is not simple to establish. The primary method used to address the question is ``brute force'' resolution study, in which the dependence of results on the relevant parameters is studied numerically. The limitation of this approach is that in practice the largest feasible simulation always sets the benchmark for what is converged.
More insidiously, there is an implicit assumption that this numerical convergence also corresponds to a convergence to the true physical result. 

Despite the impressive size of current simulations, that this remains a real practical 
concern is illustrated by the fact that the conclusions drawn from such studies continue to evolve in time. For example, the widely used ``halofit" model of \cite{smith2003stable} for power spectra were shown by higher resolution studies  
\citep{takahashi_etal_2012, benhaiem2013self, benhaiem_etal_2017} to very significantly 
underestimate power at small scales.  A more recent example is the extensive convergence study of \cite{euclid_collab_2019a}, carried out in the context of the preparation of the Euclid mission, which concludes, in agreement with another recent study of \cite{klypin+prada_2019}, that the dimension of the smallest simulation box needed to attain one  $1\%$ accuracy in the power spectrum for $k<1$ hMpc$^{-1}$ is $1250$ h$^{-1}$Mpc, while an analogous large previous study by \cite{schneider_et_al_2015}, for which the maximal size was slightly smaller, concluded that it would suffice to have a box of dimension $500$ h$^{-1}$Mpc.  Another method to test simulation convergence which has been used recently in the same context is the comparison of different codes (see e.g. \cite{schneider_et_al_2015}). The problem with this method is that it can only give confidence that codes are correctly calculating the clustering in the $N$-body system, while begging the essential question of the effect of the $N$-body discretisation. 

In summary, despite numerous studies over the last two decades we do not have a
method for determining reliably the precision even of two-point statistics calculated in $N$-body simulations. The methods which are currently used suffer, as we have underlined,  from the limitation that they can establish {\it relative} convergence, i.e., proximity to a result that must itself be assumed to be correct. In this paper we show that an approach to the problem using a specific property of scale-free models --- with an initial power-law power spectrum of fluctuations and an Einstein de Sitter expansion law --- appear to provide a more reliable manner of establishing convergence and quantifying precision. These models are well known since early studies by Peebles \citep{peebles}, who 
highlighted their self-similarity and derived analytical predictions for them in the 
so-called stable clustering approximation. They have been quite extensively studied using $N$-body 
simulations. Such studies have focused either on their use in evaluating this stable clustering approximation  \citep{efstathiou1988gravitational, padmanabhan1995pattern, colombi_etal_1996, bertschinger_98,jain+bertschinger_1998, valageas_etal_2000, smith2003stable, widrow_etal2009,
benhaiem2013self, benhaiem_etal_2017}, or have exploited them as a simplified class of models on which to calibrate numerical ansatzes for cosmological models, either for two-point 
properties (see e.g. \cite{peacock, smith2003stable}) or halo properties (see e.g. 
\cite{Navarro_etal_1997, cole1996structure, knollmann_etal_2008, diemer+kravtsov_2015, ludlow+angulo_2017, diemer+joyce_2019}). 

Testing of the $N$-body method using scale-free models has in practice, however, almost exclusively been limited to a qualitative analysis, other than in some recent studies \citep{orban2013keeping, benhaiem2013self, benhaiem_etal_2017} that underline and show the possibility of obtaining {\it quantitative} information about resolution using scale-free models, 
We apply and develop here the kind of approach used by \cite{benhaiem2013self, benhaiem_etal_2017}.
We show, using simulations of the size of current typical large cosmological simulations performed with the \Abacus code \citep{garrison_et_al_2018,garrison_et_al_2019}, that we can indeed obtain precise information about convergence in this way, and even a determination of spatial 
resolution as a function of time at levels of precision which are as stringent at those required in the context of forthcoming observational programs.

The article is structured as follows. In the next section we explain the self-similar evolution of scale-free models
and how it provides a method in principle for determining the precision with which any given statistical quantity is measured in an $N$-body simulation. 
In Section \ref{section-Abacus+sim} we describe the \Abacus $N$-body code and then specify the parameters characterizing the set of scale-free simulations we perform.  In Section 
\ref{section-time-stepping} we consider carefully the convergence of our chosen $N$-body system with respect to its numerical parameters, and in particular the parameter controlling time-stepping. 
In Section \ref{results-resolved-scales} we then present our main results and analysis methods, showing in detail how we obtain, using the constraint of self-similarity, a determination of the scales in which the physical limit is resolved to an estimated precision. In Section \ref{Interpretation} we describe our physical interpretation of these results and in particular show that the late time evolution of the lower cut-off to resolution agrees well with a simple analytical estimate derived assuming that it arises from
two-body collisionality. In Section \ref{LCDM} we show how with simple and reasonable assumptions we can use our results to estimate also the resolution limits of LCDM simulations. We conclude with a discussion of our results in relation to the existing literature and outline how we intend to develop our study of related issues in forthcoming work.
 
\section{Resolution of $N$-body simulations and self-similarity of scale-free models} 
\label{Resolution-NBS} 

The physical inputs to an $N$-body simulation of a cosmological model are its linear power spectrum $P(k)$ and the parameters fixing the evolution of the background. Setting up the $N$-body configuration and specifying the $N$-body dynamics involves necessarily the choice of three {\it unphysical} length scales: the mean interparticle spacing (which we will denote by $\Lambda$), the force softening scale (denoted 
$\epsilon$), and the side of the periodic box (denoted $L$). Finally it requires also a specification of the 
starting red-shift $z_i$, which can be conveniently parametrized by $\sigma_i$, the square root of the
variance of normalized linear mass fluctuations in a top-hat sphere of radius $\Lambda$ (with $z_i$ being chosen so that this quantity is small). Depending on the code, there may also be further parameters: for example,  $\epsilon$ itself may evolve as a function of time or the particle sampling itself may be modified (in codes implementing refinements). We will refer to these unphysical parameters as the {\it discretization parameters}. Solving the $N$-body problem numerically introduces further parameters controlling the approximations made in calculating the forces and integrating the dynamics, which we will refer to as the {\it numerical parameters}.  

The issue of convergence of $N$-body simulations and their resolution breaks then naturally into two different questions. On the one hand, the convergence of the numerical solution of the well specified $N$-body problem at {\it fixed} values of $\Lambda$, $\epsilon$, $L$, and $\sigma_i$.
On the other hand, the convergence  of the quantities calculated from the $N$-body configurations 
at given discretization parameters to those in the underlying continuum cosmological model, obtained in principle by an appropriate extrapolation of the discretization parameters (with $\Lambda \rightarrow 0$, $\epsilon \rightarrow 0$, $L \rightarrow \infty$ and $\sigma_i \rightarrow 0$). 
The first question is in principle straightforward and can be treated by studying the stability of results under variation of the numerical parameters controlling time stepping and the accuracy of force calculations. The second question is the more complex one we are primarily concerned with here
and which is currently lacking a satisfactory answer. This is true in particular for what concerns the convergence with respect to the parameters $\Lambda$ and $\epsilon$, about which there is even qualitative disagreement in the literature (which we will discuss in the final section below).   

The relevance of scale-free models in this context is the following. The question of physical resolution of $N$-body simulations is in practice that of the dependence of results on unphysical length (or time or mass) scales. Let us suppose now that the cosmological model simulated is a scale-free model, i.e.~the input linear power spectrum is taken to be a pure power-law $P(k) \propto k^n$, and the input cosmology assumed to 
be Einstein-de-Sitter with expansion rate $H^2  \propto a^{-3}$ (and thus $a \propto t^{2/3}$, where $a$ is the scale factor). In the limit that results of the corresponding $N$-body simulation do not depend on any of the unphysical parameters, there is then only one length scale and one time scale in the problem. The former
is given by the normalisation of $P(k)$ which can be taken as the non-linearity scale $R_{NL}$ 
defined by
\begin{equation}
\sigma_{\rm lin}^2 (R_{\rm NL}, a)=1
\label{RNL-def}
\end{equation}
where $\sigma_{\rm lin}^2 (r,a)$ is the variance of the normalized linear mass fluctuations in a sphere of radius $r$. The single time scale is fixed by the normalisation of the Hubble law, i.e., by Newton's constant $G$ and the initial mass density. It follows that for any  
dimensionless function $f$ characterising the clustering in the system as a function of
spatial scales $x_i$ we may write
\begin{equation}
f(x_1,x_2, \cdots,a)=f(x_1/R_0,x_2/R_0,\cdots, a/a_0)
\end{equation}
where $a_0$ is the reference scale factor and $R_0$ the value of $R_{\rm NL}$ at this time. 
However, given that the choice of this reference time is itself arbitrary, we can always choose
$a_0 =a$ as our reference,  and thus obtain
\begin{equation}
f(x_1,x_2, \cdots,a)=f_0 (x_1/R_{\rm NL} (a), x_2/R_{\rm NL} (a), \cdots )
\end{equation}
where $f_0$ is independent of time. This expresses the self-similarity of the evolution: {\it temporal evolution is equivalent to a rescaling of the spatial coordinates}. Further we have that linear theory itself furnishes 
the functional form of the evolution: from Eq.~(\ref{RNL-def}) it follows that
\begin{equation}
R_{NL} \propto a^\frac{2}{3+n}\,.
\end{equation}

Thus, for scale-free models, to the extent that the results do not depend on the unphysical simulation parameters, the clustering must be self-similar. This provides an absolute calibration in these models
for physical results obtained from simulations: to the degree that they are self-similar, we can infer that 
they represent the desired physical limit. Further, by studying carefully deviation from self-similarity, we can
in principle infer precise information about how close results obtained are to the physical result. Indeed such deviations manifest themselves as dependence on the unphysical parameters, which allow us to infer how resolution depends on them. We note this self-similar behaviour must apply to any physical quantity and thus the method can be used to evaluate the resolution of simulations for any statistic, with the form of the self-similar transformation easily inferred by writing an appropriate dimensionless form. For example for the halo mass function $n(M,a)$ it is convenient to define a characteristic non-linear mass scale 
$M_{\rm NL} \propto R_{\rm NL}^3$ and self-similar scaling corresponds to the combination 
 $M_{\rm NL}^2 n (M,a)$ being time invariant when expressed as a function of the dimensionless variable $M/M_{\rm NL}$.

In this paper our goal is to develop this method for determining resolution and demonstrate its usefulness.
To do so we will focus on just one statistic, the two-point correlation function (2PCF), and consider a single scale-free model, choosing $n=-2$ as this value is close to the logarithmic slope of the $LCDM$ power spectrum in a range of scales of moderately non-linear clustering at low red-shift. We will discuss briefly in our conclusions how we envisage, in future work, both to refine the method using simulations of a class of scale-free models to determine very precisely the resolution in a class of $LCDM$ models, and to apply the method to other statistics.

\section{Numerical Simulations} 
\label{section-Abacus+sim}
In this section we give some background on the \Abacus code and the setting up of initial conditions.
We then detail our choice of simulation parameters. 

\subsection{The \Abacus $N$-body code}
\label{ABACUS}

\Abacus is a code for large cosmological $N$-body simulations based on an exact decomposition of the near-field and far-field gravitational force.  The near-field is solved with direct pair-wise summation, and the far-field is solved with a high-order multipole method.  The exact nature of the decomposition means that high force accuracy may be achieved at relatively low computational cost.  The near-field force computation is accelerated with GPUs, and the far-field is computed with a convolution over multipoles and is thus efficient in Fourier space.  The mathematical method of the force decomposition was originally developed in \cite{metchnik_2009}; the \Abacus code has been presented  in \cite{garrison_et_al_2018}
and  \cite{garrison_et_al_2019}, and will be more fully detailed in forthcoming publications (Garrison et al.~(in prep.), Pinto et al.~(in prep.) for the  modern version of the mathematical method).

The \Abacus domain decomposition is a simple cubic grid of such cells with a mean particle occupation of a few dozen and is static throughout the course of a simulation.  The decomposition of the force solver is based on this grid.  Multipoles are computed about cell centers, with the multipole order controlling the accuracy of the expansion.  We employ order 8 in this work, which is our standard choice for simulations. We consider order 8 conservative; the resulting median fractional force error is about $10^{-5}$ \citep{garrison_et_al_2019}.  The number of cells is a performance tuning parameter and has no impact on the accuracy for practical purposes.

The accuracy of \Abacus's forces is crucial for this work.  With only one parameter controlling the accuracy of the entire force solver (the multipole order), we can focus on variations of the primary remaining numerical parameter---the time step---in addition to the discretization parameters common to all N-body codes.

We employ compact spline force softening in this work. The traditional Plummer softening \citep{plummer_1911}, with the form $\mathbf{F}(\mathbf{r}) = \mathbf{r}/(r^2 + \epsilon^2)^{3/2}$, results in suppression of power on scales many times larger than the softening scale $\epsilon$ \citep{garrison_et_al_2016,garrison_et_al_2019}.  Instead 
we adopt \citep{garrison_et_al_2016} a spline softening from a Taylor expansion 
of Plummer softening in $r$, requiring continuity up to the second derivative in the transition to $1/r^2$ at the spline radius $\epsilon_s$.  The resulting force law is
\begin{equation}
\mathbf{F}(\mathbf{r}) =
\begin{cases}
\left[10 - 15(r/\epsilon_s) + 6(r/\epsilon_s)^2\right]\mathbf{r}/\epsilon_s^3, & r < \epsilon_s; \\
\mathbf{r}/r^3, & r \ge \epsilon_s.
\end{cases}
\end{equation}
To ease comparison with the literature, we quote our softening length as the Plummer-equivalent value $\epsilon$, using  $\epsilon_s = 2.16\epsilon$ as this is the value for which the
two-body orbital times of the force laws agree at small radius.

Aside from the softening length and time step,  which we discuss next, the only remaining parameter controlling the simulation accuracy is the multipole order.  We employ order 8 in this work, which is our standard choice for simulations. We consider order 8 conservative; the resulting median fractional force error is about $10^{-5}$ \citep{garrison_et_al_2019}.  The number of cells is a performance tuning parameter and has no impact on the accuracy for practical purposes.

\subsection{Time stepping}\label{sec:timestep}
\Abacus is currently a globally time-stepped code, in which all particles share the same time step.  This is a simple and accurate scheme, since the tightest orbit (usually in the center of a massive cluster) sets the time step for the entire simulation. While this is computationally expensive, due to the wide range of dynamical times in the simulation,  it has the benefit of minimizing integration errors in the particle dynamics.

The time step $\Delta a$ is determined, at each integration step,  as the smaller  of two possible time steps. 
The first is computed from the instantaneous particle velocities and accelerations as 
\begin{equation}
\label{eqn:timestep}
\Delta a = \eta\max\left(\min_\mathrm{c \in cells}\left[\frac{v_\mathrm{rms,c}}{a_\mathrm{max,c}}\right], \frac{v_\mathrm{rms,global}}{a_\mathrm{max,global}}\right)\,,
\end{equation}
where $\eta$ is an input parameter (\texttt{TimeStepAccel}  in the code) and the cells are those of the \Abacus domain decomposition (see above). The minimum here is taken over the ratio of RMS velocity to maximum acceleration computed within each cell (in suitable units). This minimum is compared with the ratio of global $v_\mathrm{rms}$ to global $a_\mathrm{max}$, and the maximum of the two is taken.  This guards against taking catastrophically small time steps as a result of abnormally cold cells.  Since the most demanding $v_\mathrm{rms}/a_\mathrm{max}$ requirement in the whole box sets the time step, we expect that an insufficient time step will most readily reveal itself in the cores of the most massive halos. This value of $\Delta a$ is then adopted as the time step unless it is larger than a chosen constant time-stepping in $\log a$, specified  by an input parameter
$\eta_{\rm H}$ (\texttt{TimeStepDlna}  in the code).

In past work  \citep{garrison_et_al_2019}  we have found good stability of results, e.g. at order 1\% in the matter correlation function on scales close to the softening length, with 
$\eta$ between $0.15$ and $0.3$, and values of $\eta_{\rm H}$  for which the time-step is given by Eq.~(\ref{eqn:timestep}) except at very early times when the displacements from the lattice are still small. Indeed once $\eta_{\rm H}$ is chosen sufficiently small ($\ltapprox 0.03$) so that these early epochs, are integrated at high accuracy it is solely the parameter $\eta$ which controls the accuracy of results at a relevant level.  In this work we explore very carefully  the convergence of our simulations with respect to time-stepping, by varying the parameter $\eta$  with fixed $\eta_{\rm H}=0.03$.   We note that our time-stepping  is thus not a priori constant in $\log a$ other than at very early times, but that the integration of scale-free models with such a constant time-stepping would in fact be interesting to explore:  in this case the time discretized dynamics introduce no new time-scale and  thus  admit self-similar solutions distinct from that of the continuum model\footnote{We thank A. Jenkins for pointing this out to us.}. We will discuss the point a little more at the end of Section \ref{section-time-stepping}. 

\subsection{Initial conditions}
\label{Initial conditions}

We set up our scale-free Gaussian initial conditions (ICs) as described in \cite{garrison_et_al_2016}.  The method begins with the standard procedure of generating a random realization of the power spectrum amplitudes and phases on a mesh, but differs in how particle displacements are determined from these modes.  Rather than assuming that the growing mode is the curl-free solution of a continuous density field, the method instead uses the growing mode of the particle system, which unavoidably contains a curl component.  This ``particle linear theory'' (PLT) introduced in \cite{joyce_et_al_PLT_2005} (see also \cite{marcos_et_al_PLT_2006,  discreteness2_mjbm}) enables prediction of how the particle system's growth will deviate from the prediction of continuum linear theory.  We can thus correct for this deviation: initial mode amplitudes are modulated as a function of wave-vector such that the system will arrive at the prediction of linear theory at a target scale-factor $a_{\rm PLT}$. Its choice for this work will be specified below. A more thorough examination of the effect of initial conditions on non-linear structures using scale-free simulations will be conducted in future work.

We generate 2LPT corrections using the configuration-space method of \cite{garrison_et_al_2016}: the sum of two force calculations with opposing particle displacements yields the 2LPT correction.  Since we use direct force evaluations, this is the correct 2LPT for the dynamical system represented by particles interacting via \Abacus forces.  The 2LPT corrections are generated via modified Kick and Drift operators as the first two time steps of each simulation.

\subsection{Simulations parameters and outputs}

We consider here an $n=-2$ scale-free cosmology, with a standard Einstein-de-Sitter expansion rate 
(i.e. $H^2=8\pi G \rho/3$ where $\rho$ is the mass density).  We choose this value of $n$
as it is close to the logarithmic slope of the power spectrum of LCDM models
at physical scales probed in typical cosmological simulations, while also, as we will see, also being sufficiently
far from $n=-3$ that finite box size effects remain sufficiently small for the particle 
numbers we simulate.

Given that we will be focusing in particular on the limits on resolution at small scales,
it is convenient to use $\Lambda$ to define our units of length. To specify fully
the $N$-body system we then need just the number of  particles $N$ (or $L=N^{1/3}$), 
the ratio $\epsilon/\Lambda$, and, for the IC, the parameters $a_{\rm PLT}$ and $\sigma_i$. 

We report here results of simulations of the $N$-body system with  $N=1024^3$ 
and $\epsilon/\Lambda=1/30$ (a typical value of the 
force smoothing employed in cosmological simulations). For the amplitude
of the initial conditions at the starting time, we take $\sigma_i=0.03$, which is 
sufficiently small so that we expect no relevant improvement in accuracy can be attained by starting at lower amplitude. 
It is useful to define a reference scale-factor $a_0$ by
\begin{equation}
\sigma_{\rm lin} (\Lambda, a_0)= 0.56 
\label{a0-ref}
\end{equation}
This corresponds to the time at which we expect fluctuations with values $\nu \approx 3$ of the
standard peak height parameter to virialize, according to the standard estimate from the spherical 
collapse model ( $\nu=\delta_c/\sigma_{\rm lin}$ and  $\delta_c=1.68$). It gives an estimate of the time at
which the very first non-linear structures appear in the simulation. Given that it is around such a 
time that we wish the linearly evolved fluctuations to most faithfully represent the continuum model,
we also adopt it as the target scale factor $a_{\rm PLT}$ for our PLT correction i.e. we take
$a_{\rm PLT}=a_0$. As noted we will study directly 
the effects of varying all these parameters in a subsequent article. Here our goal is to show that we 
can determine well the resolution limits imposed by these choices defining the $N$-body system by using self-similarity alone.

We run our simulations of this system up to a scale-factor $a_f$ 
which is determined in practice by the numerical cost of the integration. 
In \Abacus we can integrate efficiently (given an appropriate time stepping criterion, 
as described above) until the time at which  
$\sigma_{\rm lin} (100 \Lambda) \approx 0.5$, 
i.e.  up to a time at which the typical virializing halos have of order one million particles.  
Irrespective of such numerical limitations, self-similarity will necessarily be broken 
strongly at some time due to the finite box size. 
In this respect it is relevant to characterize the final configuration (at $a=a_f$) by e.g.  
\begin{equation}
\sigma_f=\sigma_{\rm lin} (\frac{L}{2}, a_f)
\end{equation}
which one expect may be a control parameter for the importance of finite size effects. 
We have stopped our simulation at $a_f/a_0\approx 8.48$, which corresponds to
$\sigma_f=0.21$. We will see below that we do indeed detect such finite size effects in our simulations well before this time.

Starting from $a_0$ as defined in Eq.~(\ref{a0-ref}), we save our outputs at a series 
of $S$ scale-factors $\{a_0,a_1,...a_{S-1}\}$ with equal logarithmic spacing 
\begin{equation}
\log_2 (\frac{a_{s+1}}{a_{s}})= \frac{1}{12} \,.
\label{interval-def}
\end{equation}
Given that  $M_{\rm NL} \propto R_{\rm NL}^3 \propto a^{6}$, this choice corresponds to
\begin{equation}
\log_2 \frac{M_{\rm NL}(a_{s+1})}{M_{\rm NL}(a_{s})}=\frac{1}{2} 
\label{mass-interval}
\end{equation}
i.e. the intervals between successive outputs are such that the non-linear mass
grows by a factor of $\sqrt{2}$ (and $R_{\rm NL}$ grows by the same factor over 
three outputs).
We will give all our results in terms of the time variable $\log_2(a/a_0)$,
where for snapshot $s$ ( $s \in \{0,S-1\}$)
\begin{equation}
\log_2(\frac{a_{s}}{a_{0}})= s/12
\label{s-loga-relation}
\end{equation}
We have $S$=$38$ snapshots, corresponding to $a_f/a_0$ just slightly larger than $8$. 
We note also that
$a_0/a_i=(0.56)/(0.03) \approx 18.7$
where $a_i$ is the scale factor at the starting time (when $\sigma_i=0.03$).
Thus if we define red-shift $z$ with $z=0$ at our final time, the starting red-shift of 
our simulation is $z_i \approx  157$, while $a=a_0$ corresponds to $z \approx 7.5$.

We calculate 2PCFs on the particle data using the publicly available code {\sc corrfunc}  \citep{Sinha_Garrison_2020} in about 40 equally spaced logarithmic bins per decade.
For convenience, given that we wish to compare 2PCFs at rescaled distances, we calculate the 2PCFs directly in appropriately rescaled bins. As the calculation of the 2PCFs on the largest simulations becomes very costly as separation increases, we calculate only up to $r \sim 10 \Lambda$, which is sufficient for our purposes.

\section{Results: Convergence of $N$-body integration} 
\label{section-time-stepping}

\begin{figure}
\centering\resizebox{8cm}{!}{\includegraphics[]{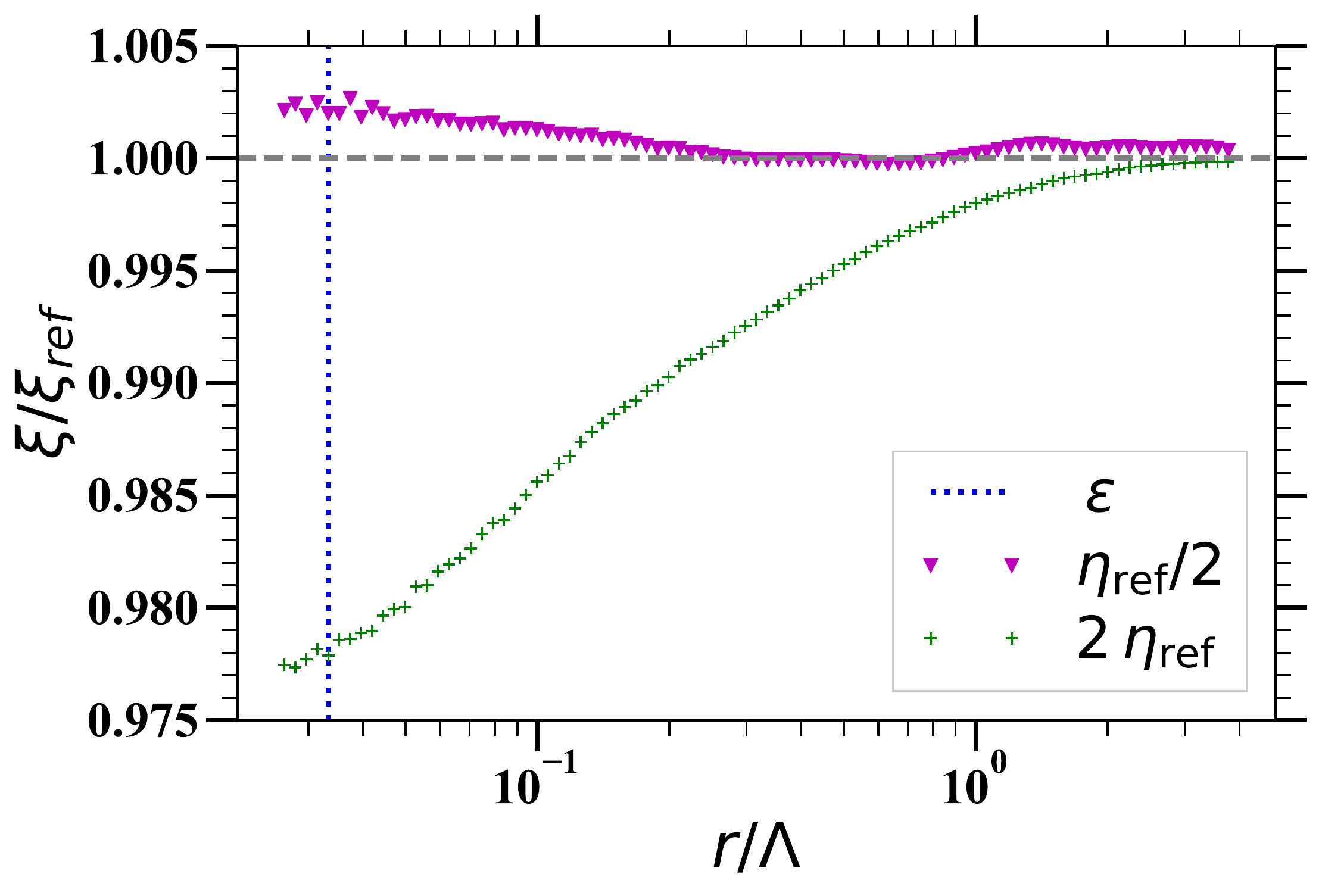}}
\caption{Ratios of the 2PCF measured in simulations with modified time stepping parameter $\eta$  to that in our reference simulation ($\eta=\eta_{\rm ref}=0.15$), at the final time ($\log_2(a/a_0)$=$3.08$)  when these ratios are maximal. Given the observed rapidity of the convergence, we infer a conservative upper bound of $0.3\%$ on the accuracy of our reference simulation.}
\label{fig-time-stepping-1}
\end{figure} 

\begin{figure}
\centering\resizebox{8cm}{!}{\includegraphics[]{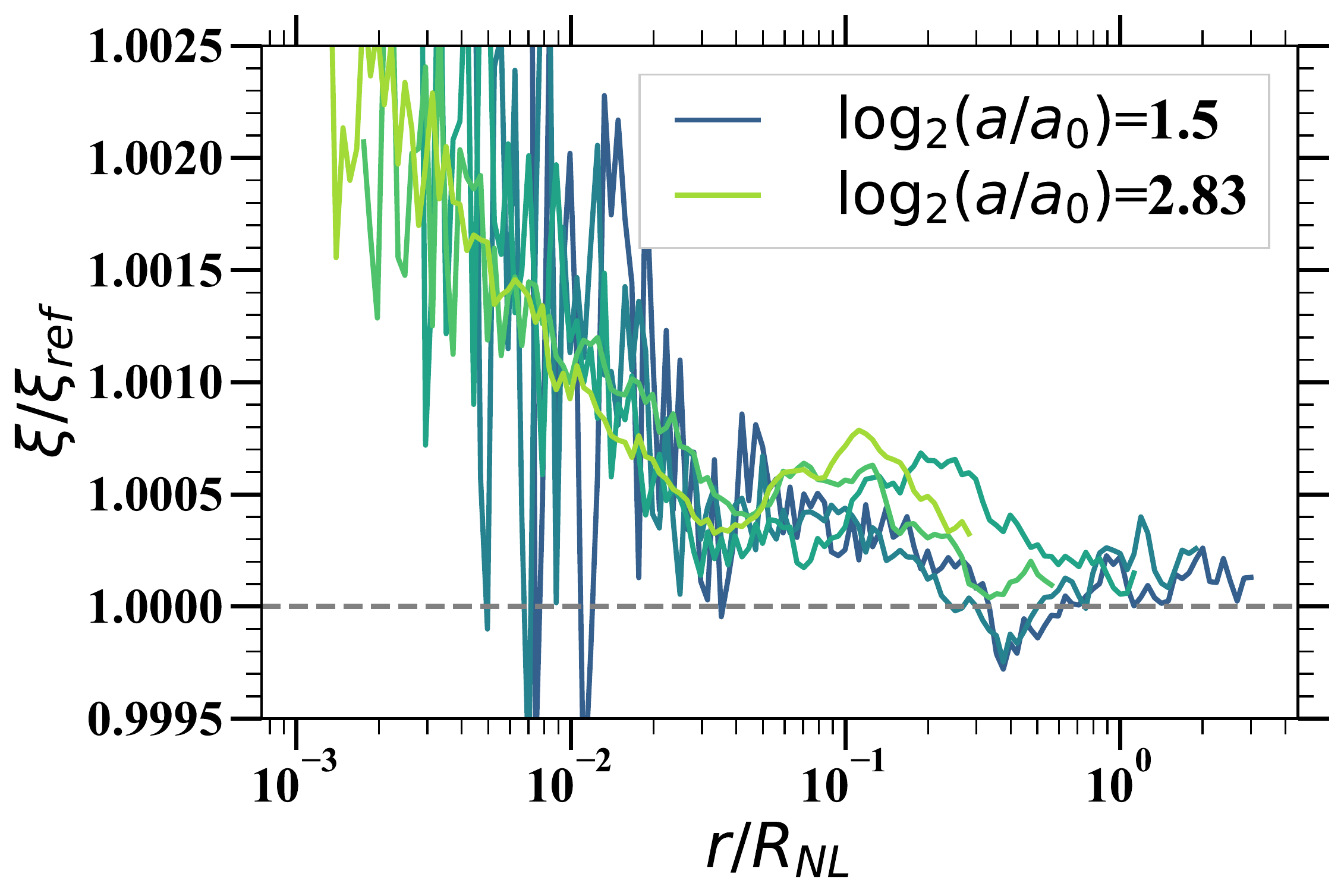}}
\caption{Ratios of the 2PCF measured in simulations with the time stepping parameter $\eta$ 
equal to half its value in our reference simulation to that in the reference simulation. Each curve 
is for a different time, with the times corresponding to equally logarithmically spaced scale factors 
between the largest and smallest values indicated in the legend; note that the $x$-axis has been rescaled by the non-linearity scale  $R_{NL}$ to obtain the approximate superposition of the curves. }
\label{fig-time-stepping-2}
\end{figure} 

Before performing our self-similarity analysis, we need to consider how well the $N$-body simulation, specified by the parameters given above, is converged with respect to the numerical parameters controlling its accuracy. More specifically we need to know what accuracy we can assume for the 2PCF in the range of scales which will be relevant for our study. In practice, as we will show that self-similarity can be observed at the percent level (or even somewhat smaller), we need to be sure that the 2PCFs are accurate down to scales 
of order $\epsilon$ well below this level.

We adopt $\eta=0.15$ as the timestep for our reference simulation.  Previous work \citep{garrison_et_al_2019} found only a 1\% change in LCDM simulations near the softening scale when doubling to $\eta=0.3$.  To further
test this, we have run our scale-free simulations for the values
$0.075,0.15, 0.3,0.45,0.6,$ and $1.0$.  We find, as anticipated, that the latter three values are not
sufficiently small to give convergence approaching the percent level.  We thus focus on the simulations with half the reference time step ($\eta=0.075$) and twice the reference time step ($\eta=0.3$).

Figure \ref{fig-time-stepping-1} shows the ratios of the 2PCF measured in these two simulations 
with $\eta=0.075$ and $\eta=0.3$ to the 2PCF measured in our reference simulation. These
quantities are found to be monotonically increasing functions of time, and the plot is 
for the latest time simulated ($s=37$, $\log_2(a/a_0)=37/12$) at which time these ratios
are largest. As the deviation from unity of the ratios of the 2PCF in pairs of simulations 
in which $\eta$ differs by a factor of two is observed to decrease strongly (by roughly an
order of magnitude), there is clear evidence for convergence of the 2PCF.
Conservatively, the 2PCF in the reference simulation can be taken to be 
converged to below $0.3 \%$ at all scales (i.e. down to $\epsilon$), and indeed to 
below $0.1 \%$ at all but the smallest scales (above $r \approx \Lambda/5$).

Figure \ref{fig-time-stepping-2} shows again the ratio of the 2PCF measured 
in the simulation with $\eta=0.075$ to that in the reference $\eta=0.15$ simulation, but for 
a range of different times.
The $x$-axis 
has been rescaled by $R_{NL}$, as we have found that by doing so we
obtain an approximate superposition of the curves. A
characteristic scale $r/R_{NL} \approx 0.2$ marks the transition between very small 
errors in the 2PCF to a larger (albeit still small) suppression of its 
value due to finite time-stepping which increases systematically at
smaller scales. This scale 
is thus clearly associated to the transition to strong non-linearity. 

An important implication of Figure \ref{fig-time-stepping-2} is
that the numerical errors due to finite time stepping show an approximate self-similar 
scaling. To the extent that this is the case, the unconverged simulations 
will thus show approximately self-similar behaviour in any range where the
converged simulation does.  As noted at the end of Section \ref{sec:timestep}
self-similar solutions  in a time discretized version of the continuum
cosmological model are indeed expected if the temporal discretization introduces 
no additional time or length scale. This is the case of our time-stepping only at very early times,
when our time-steps have a constant spacing in  $\log(a)$, but not thereafter.
Nevertheless the time-stepping determined by the criterion in
Eq. ~(\ref{eqn:timestep} ) may lead in practice to time-stepping at later times
which is close to constant spacing in  $\log(a)$ and the approximate self-similarity 
in Figure \ref{fig-time-stepping-2}  may be a consequence of this. This is 
an interesting issue which we will investigate further in future work.
We underline that the results we have obtained here  show 
that it is essential, as we have done, to first establish convergence
of raw amplitudes with respect to time-stepping, before studying, as we 
next do,  the rescaled amplitudes to test for self-similarity as a criterion to 
identify convergence to the physical continuum limit.



\begin{figure*}
\centering\resizebox{16cm}{!}{\includegraphics[]{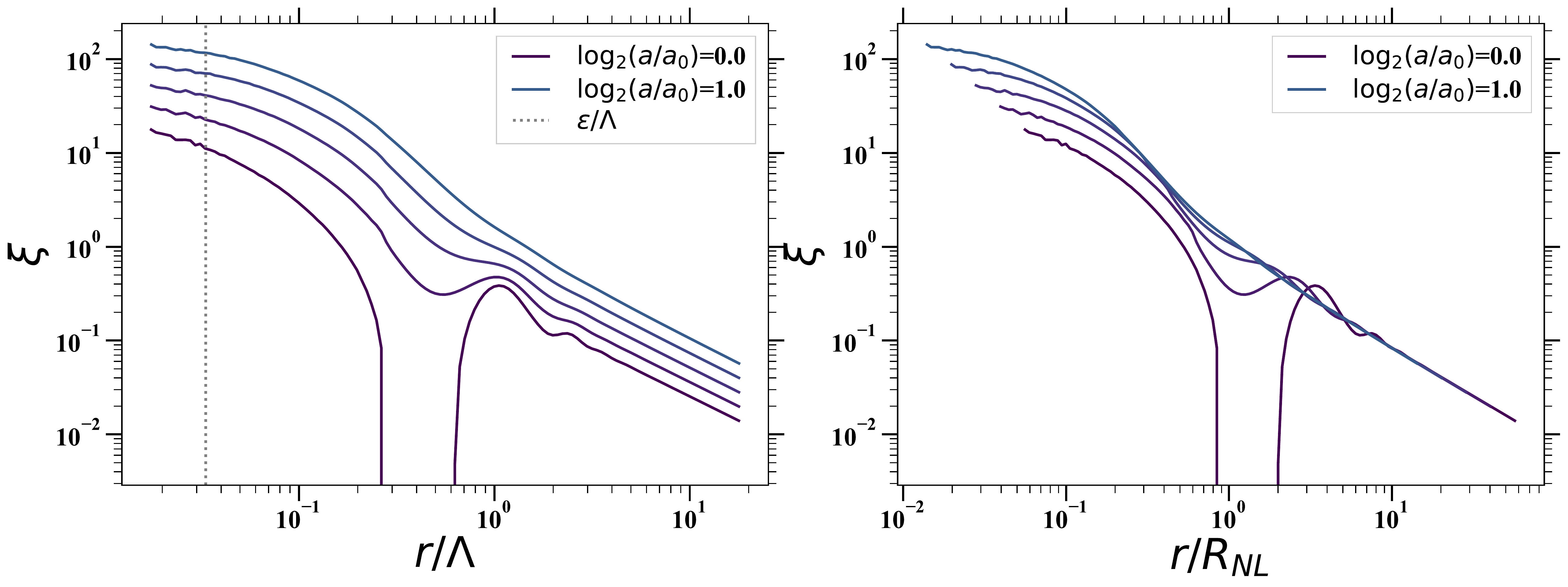}}\\
\centering\resizebox{16cm}{!}{\includegraphics[]{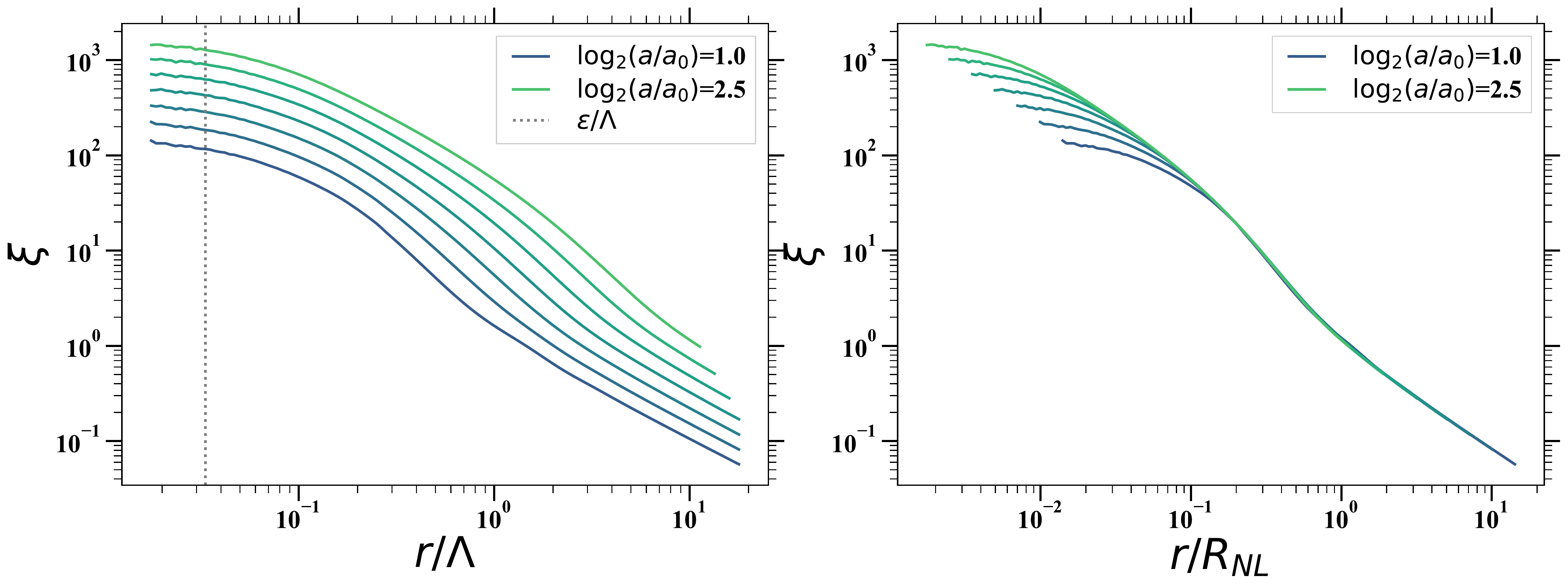}}\\
\centering\resizebox{16cm}{!}{\includegraphics[]{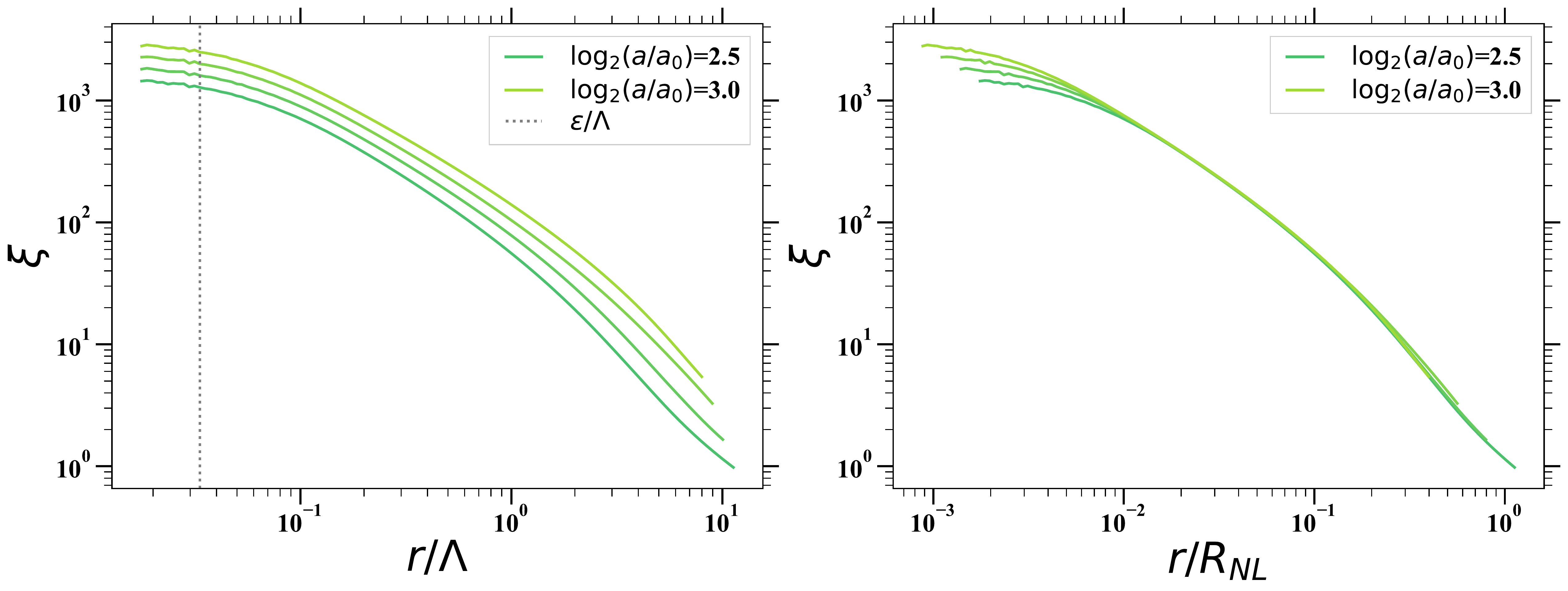}}\\
\caption{Left panels: measured 2PCFs as a function of comoving separation $r$, in units of the initial grid spacing $\Lambda$, for different phases of the evolution (equal logarithmic spacing in scale-factor between the minima and maxima indicated in the legend); Right panels: same data plotted as a function of the rescaled distance in which self-similar evolution corresponds to an invariant 2PCF.
At early time (upper panels) we see the imprint in the 2PCF of the initial grid and how it is progressively washed out; at intermediate times (middle panels) the deviation from self-similarity evolves monotonically to higher amplitudes of the 2PCF; at the latest times deviations from self-similarity due to finite box size begin to become visible.  
}
\label{fig-CFs}
\end{figure*}

\section{Results: Determination of resolved scales using self-similarity} 
\label{results-resolved-scales}

In this section we consider analysis of our reference $N$-body simulation of the $n=-2$ scale-free model
($N=1024^3$, $\epsilon=\Lambda/30$, $\sigma_i=0.03$, $a_{\rm PLT}=a_0$ and $\eta=0.15$).
We show in detail how we use the criterion of self-similarity  to infer the range of scale in which the physical 2PCF is well approximated as a function of time.

\subsection{Qualitative inspection of 2PCFs} 

Figure \ref{fig-CFs} shows the 2PCF as a function of time (specified by the scale factor $a/a_0$). We have divided the time range  into three, corresponding to an approximate division into three different phases that can be distinguished visually in these plots. In each case the panel on the left shows the 2PCFs as a function of comoving distance, in units of $\Lambda$, while the panel on the right shows the same quantities as a function of the rescaled distance i.e. divided by the scale $R_{\rm NL}$ as defined by Eq. (\ref{RNL-def}). Exact self-similarity therefore corresponds to the superposition of the curves in the right panels.  

The uppermost pair of plots in Figure \ref{fig-CFs} correspond to evolution up to $a=2a_0$. In this time range, effects from the underlying lattice are still visibly present but progressively disappear. In the middle pair of plots, corresponding to an intermediate stage of our evolution, we observe on the other hand an apparent almost perfect self-similarity down to a lower cut-off in $r/R_{\rm NL}$ which monotonically decreases with time.
The physical limit represented by the self-similar behaviour appears to be resolved  at progressively higher values of the correlation amplitude as the system evolves.
Finally, in the bottom pair of plots, corresponding to the latest times, we can observe now visible disagreements at the largest scales to which our 2PCFs extend.
These are evidently effects arising from the finite simulation box, which at these times start to affect significantly even the small scales shown in these plots.  The impact of such effects would be lesser in a LCDM simulation because the large-scale power spectrum is not as red as $n=-2$.

\subsection{Convergence and precision analysis}
\label{section-resolution}
\begin{figure*}
\centering\resizebox{16.5cm}{7.1cm}{\includegraphics[trim=0.1cm 0.1cm 0.1cm 0.1cm,clip]{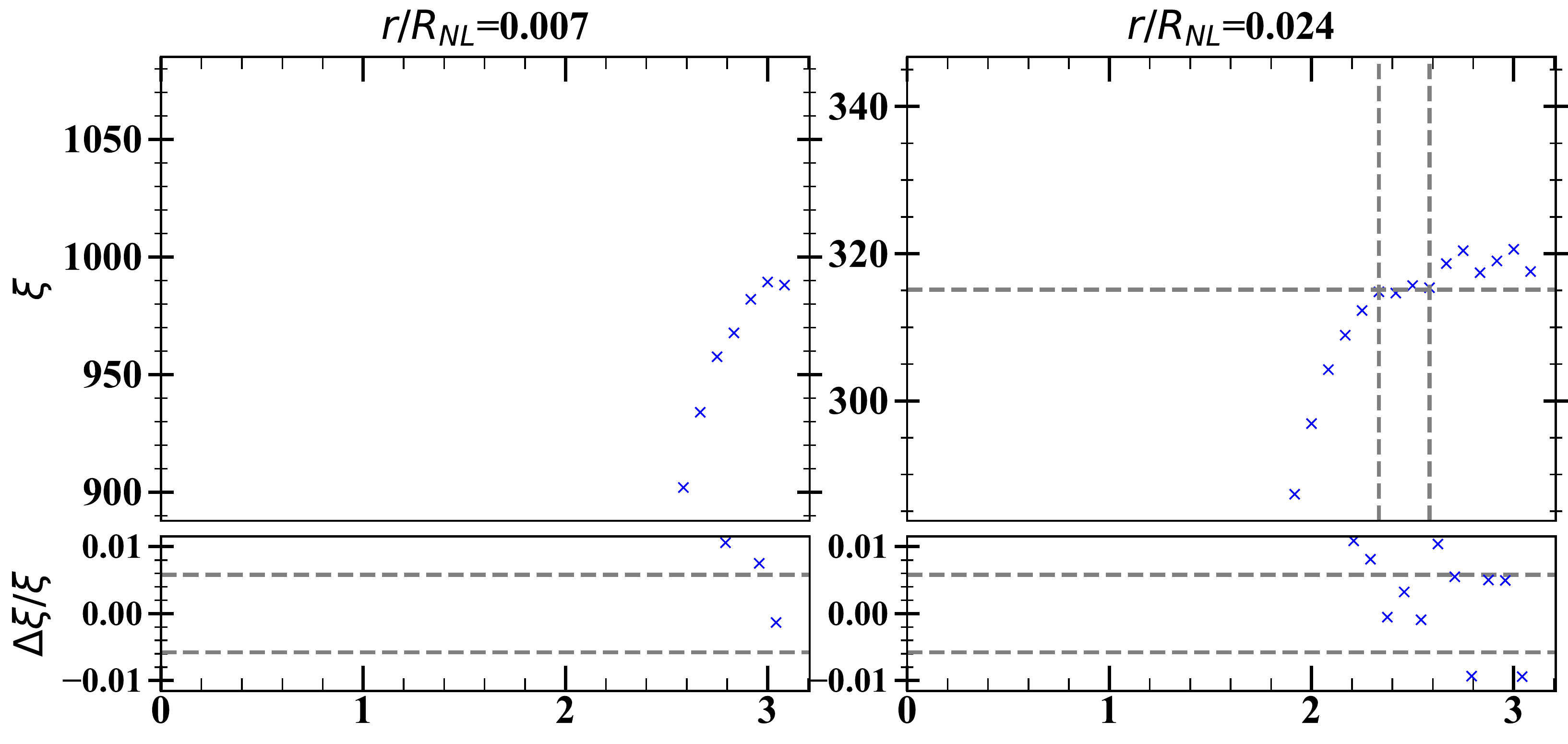}}\\
\centering\resizebox{16.5cm}{7.1cm}{\includegraphics[trim=0.1cm 0.1cm 0.1cm 0cm,clip]{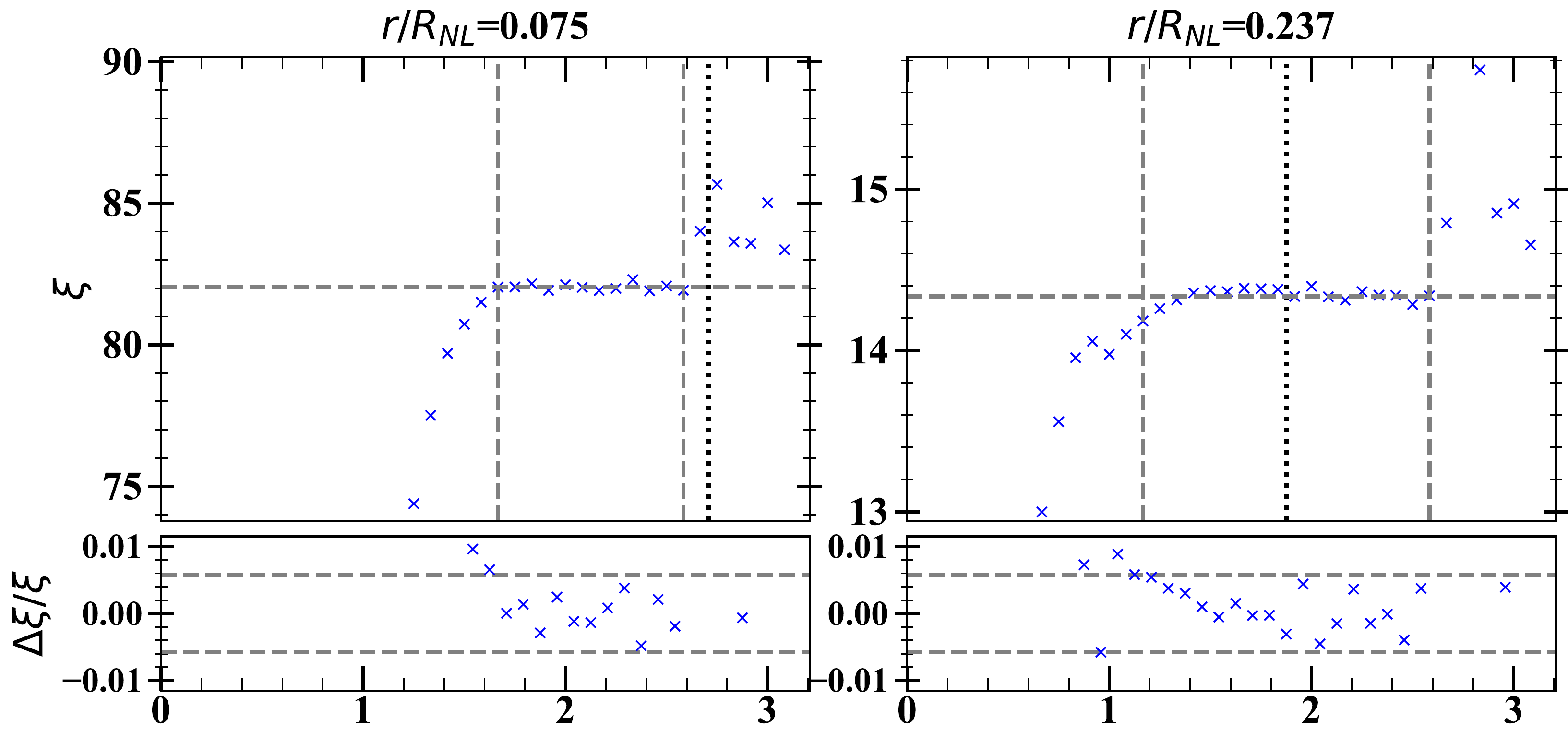}}\\
\centering\resizebox{16.5cm}{7.1cm}{\includegraphics[trim=0.1cm 0.1cm 0.1cm 0cm,clip]{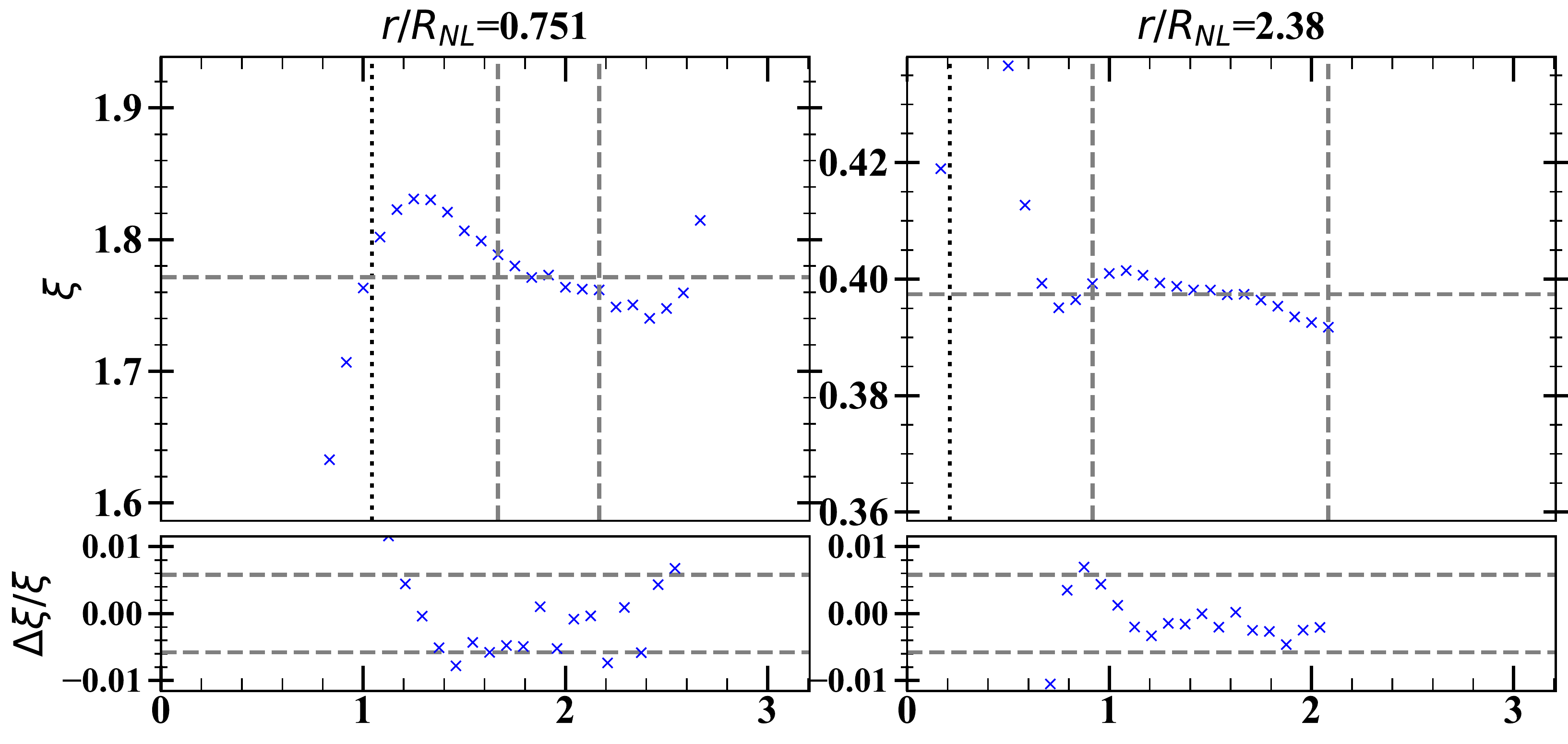}}\\
\caption{Each plot shows, for a given rescaled bin labelled by $r/R_{NL}$, the temporal evolution of the 2PCF $\xi$ (large upper panel) 
and of its fractional variation $\Delta \xi/\xi$ between adjacent snapshots (small lower panel). The vertical dashed lines in the upper panels indicate the range of scale-factors in which the convergence criterion corresponding to the horizontal dashed lines in the lower panel are satisfied: a region is considered converged if $|\Delta \xi/\xi|$ is below the chosen threshold variation for at least three consecutive snapshots. The vertical dotted lines indicates the time at which the comoving size of the rescaled bin is equal to the mean interparticle distance $\Lambda$. Starting from the second bin, we see that convergence sets 
in earlier as $r/R_{NL}$ increases, which corresponds to the behaviour 
observed visually in the middle panels of  Figure \ref{fig-CFs}. 
In the last two bins, corresponding to weakly non-linear scales,
we can see that convergence is delayed by the imprint at these
scales of the initial lattice. Deviations at later times 
due to finite box size are also apparent in all but the first bin.
} 
\label{fig-convergence}
\end{figure*}

We now turn to our quantitative analysis. Our goal is to determine, as a function of time, the range of comoving scale in which the 2PCF can be inferred to represent the physical 2PCF to some precision. 
To do so we consider the values of the 2PCF as a function of time at a fixed rescaled separation $r/R_{\rm NL}$, i.e. we consider the values of the 2PCF on a given vertical line in the right panels of Figure \ref{fig-CFs}. 
In such a representation, exact self-similarity corresponds to a time-independent value. Figure \ref{fig-convergence} shows this 
time series for the 2PCF (denoted $\xi$), as well as its
fractional variation between consecutive snapshots (denoted $\Delta \xi/\xi$),  for different chosen values of  $r/R_{\rm NL}$ spanning the relevant range. These plots show more quantitatively the trends which
have been noted in our discussion of Figure \ref{fig-CFs}: while at the smallest $r/R_{\rm NL}$ a converged self-similar behaviour is never attained, as we move to progressively larger $r/R_{\rm NL}$ the plots show a clear tendency to converge to a well defined value, which sets in earlier and becomes more and more stable as $r/R_{\rm NL}$ increases. Further in all scales showing such a convergent plateau we observe a marked, and quite abrupt, break from this behaviour close to $\log_2 (a/a_0)=2.5$. As anticipated this can clearly be ascribed to the finite box size: at this time the linear theory amplitude in a sphere of radius $L/2$ is approximately $0.15$ and structures of order the box size start to evolve non-linearly in a way which completely breaks the self-similarity. 

\begin{figure*}
\centering\resizebox{8.6cm}{!}{\includegraphics[]{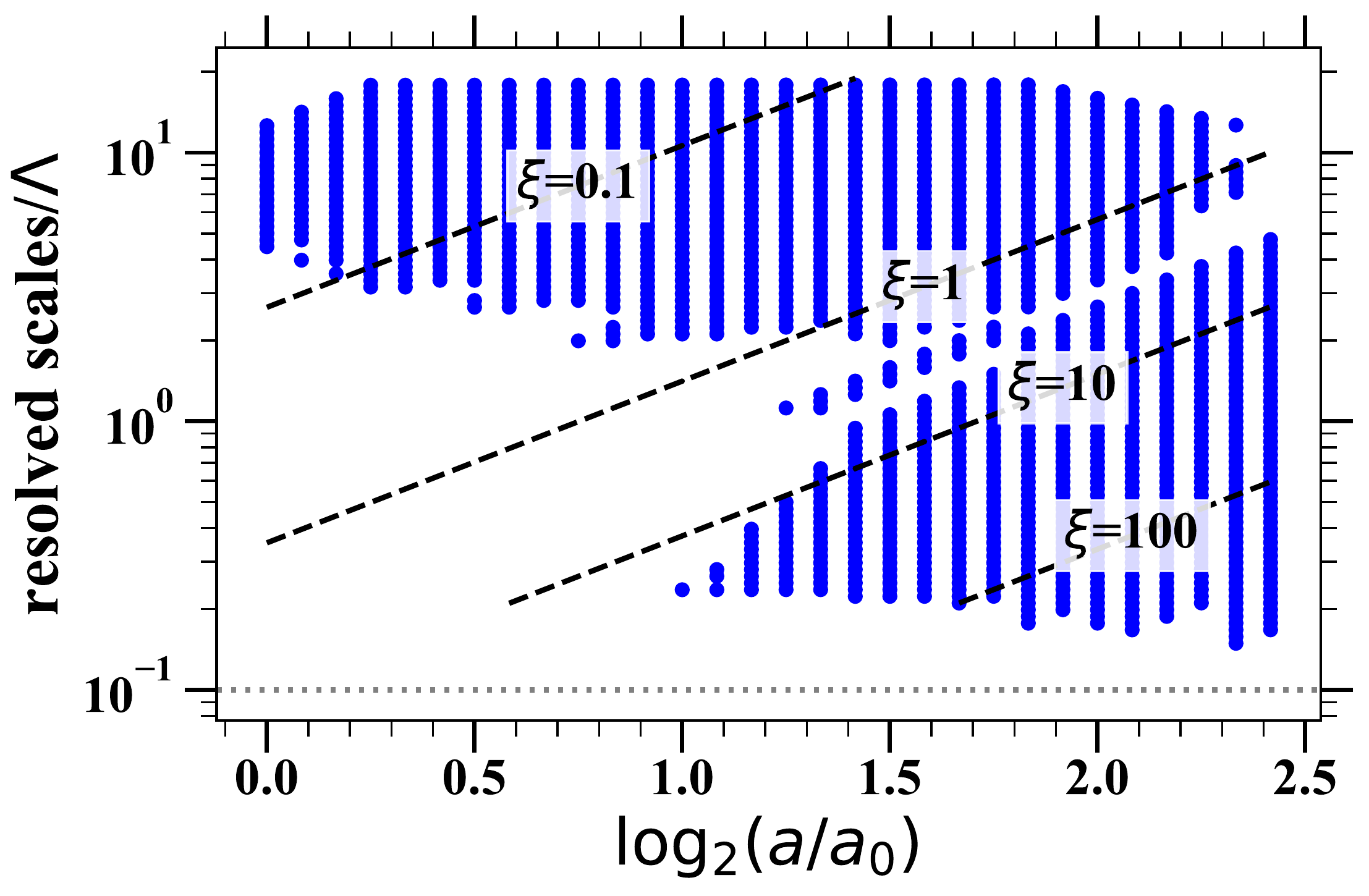}}
\centering\resizebox{8.2cm}{!}{\includegraphics[]{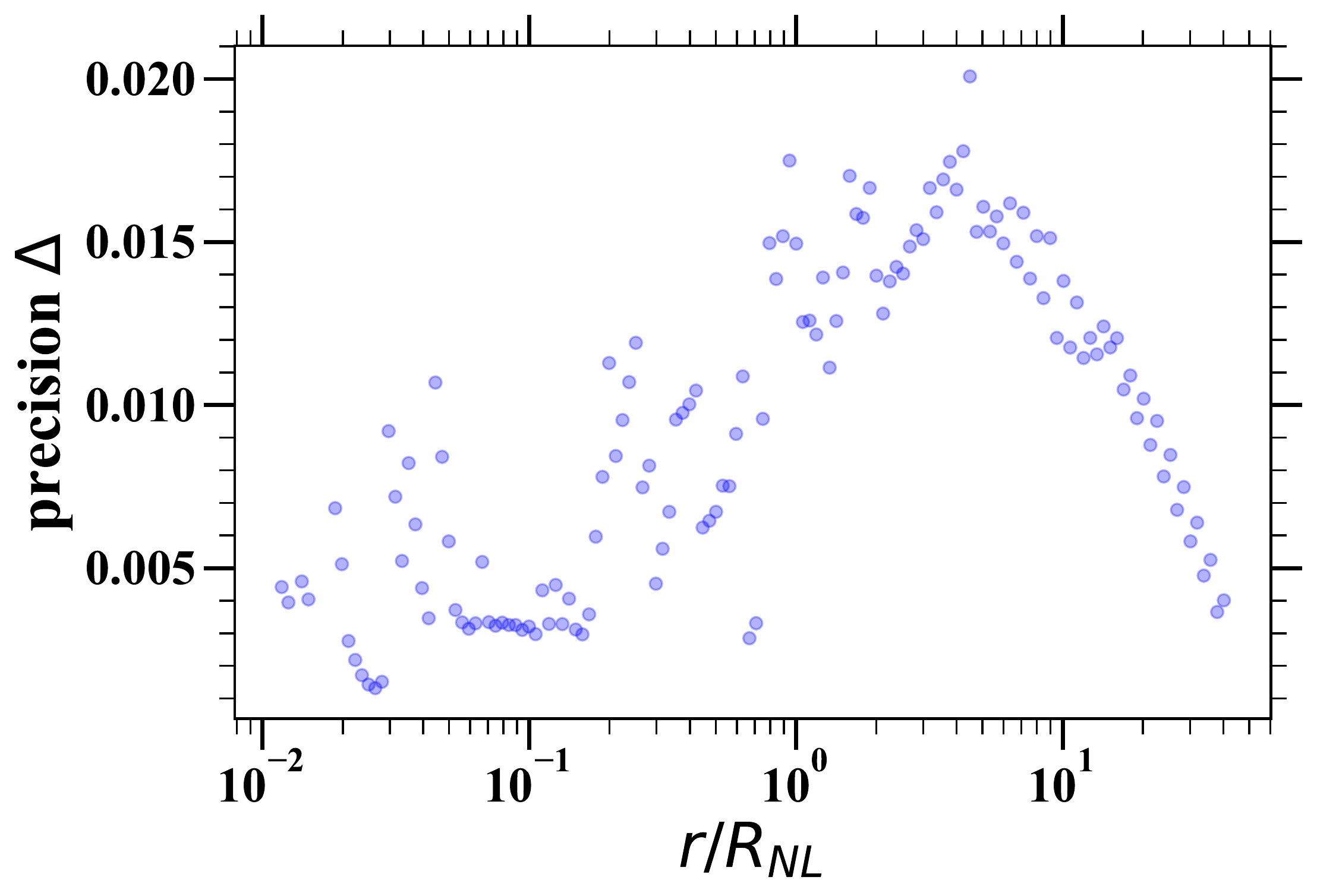}}
\caption{
Left panel: Resolved comoving length scales in units of the initial grid spacing $\Lambda$ as a function of time.; the dotted vertical line corresponds  to $3\epsilon/\Lambda$. Right panel: the associated precision of the estimated  2PCF as a function of the rescaled distance $r/R_{NL}$. The plots
correspond to the resolved regions shown by the vertical dashed lines in 
each panel of Figure \ref{fig-convergence}, selected by the criterion that
at least three consecutive snapshots have a fractional variation $\Delta \xi/\xi$
within the bounds given by the horizontal dashed lines in the subpanels
($d(\log \xi)/d(\log a) \ltapprox 0.1$). The precision $\Delta$ is the maximal value of the absolute value of the fractional difference between any value and the mean 2PCF in the resolved region
for each bin.}
\label{resolution-plots-1}
\end{figure*}

\begin{figure*}
\centering\resizebox{8.6cm}{!}{\includegraphics[]{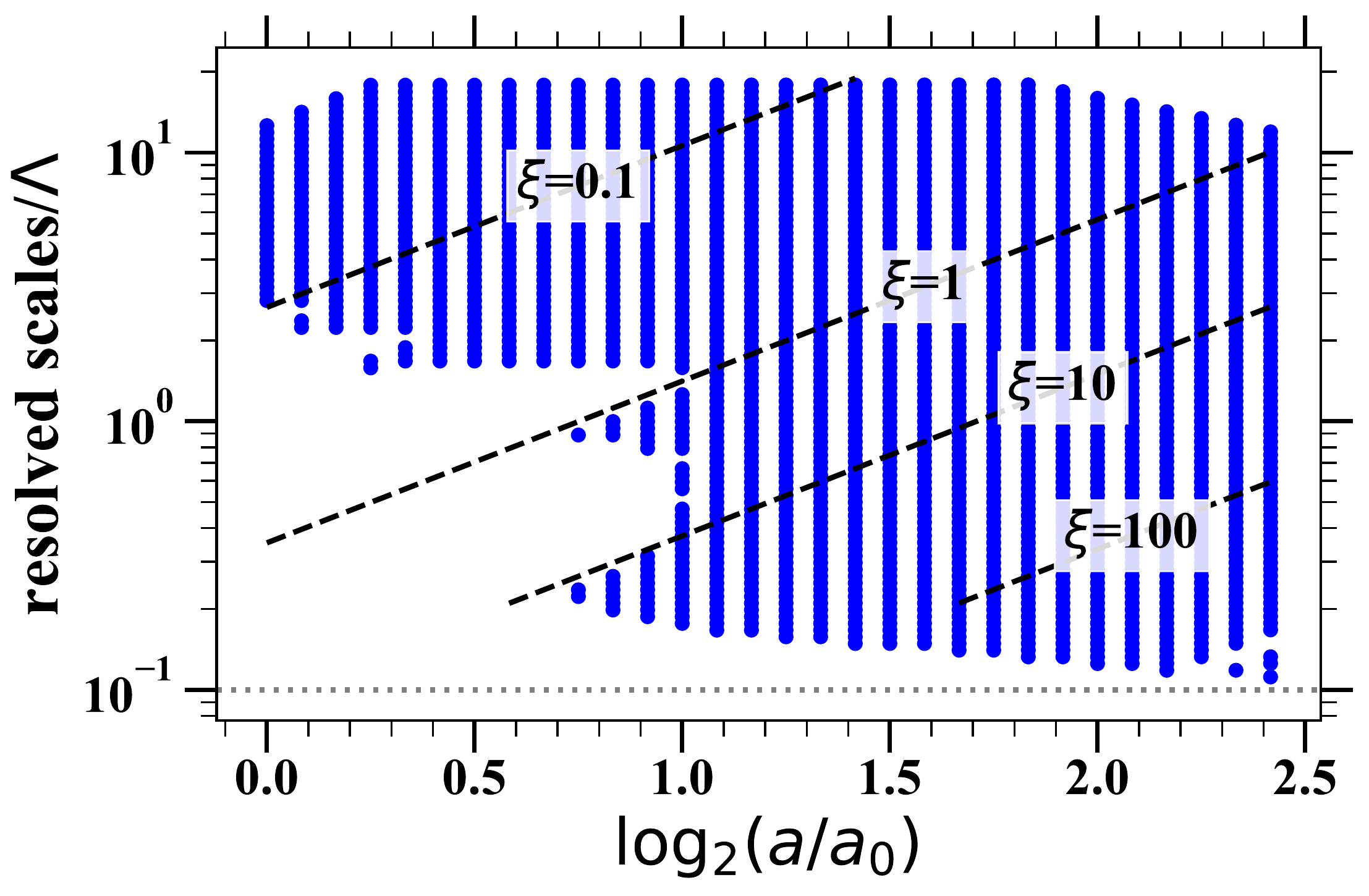}} 
\centering\resizebox{8.2cm}{!}{\includegraphics[]{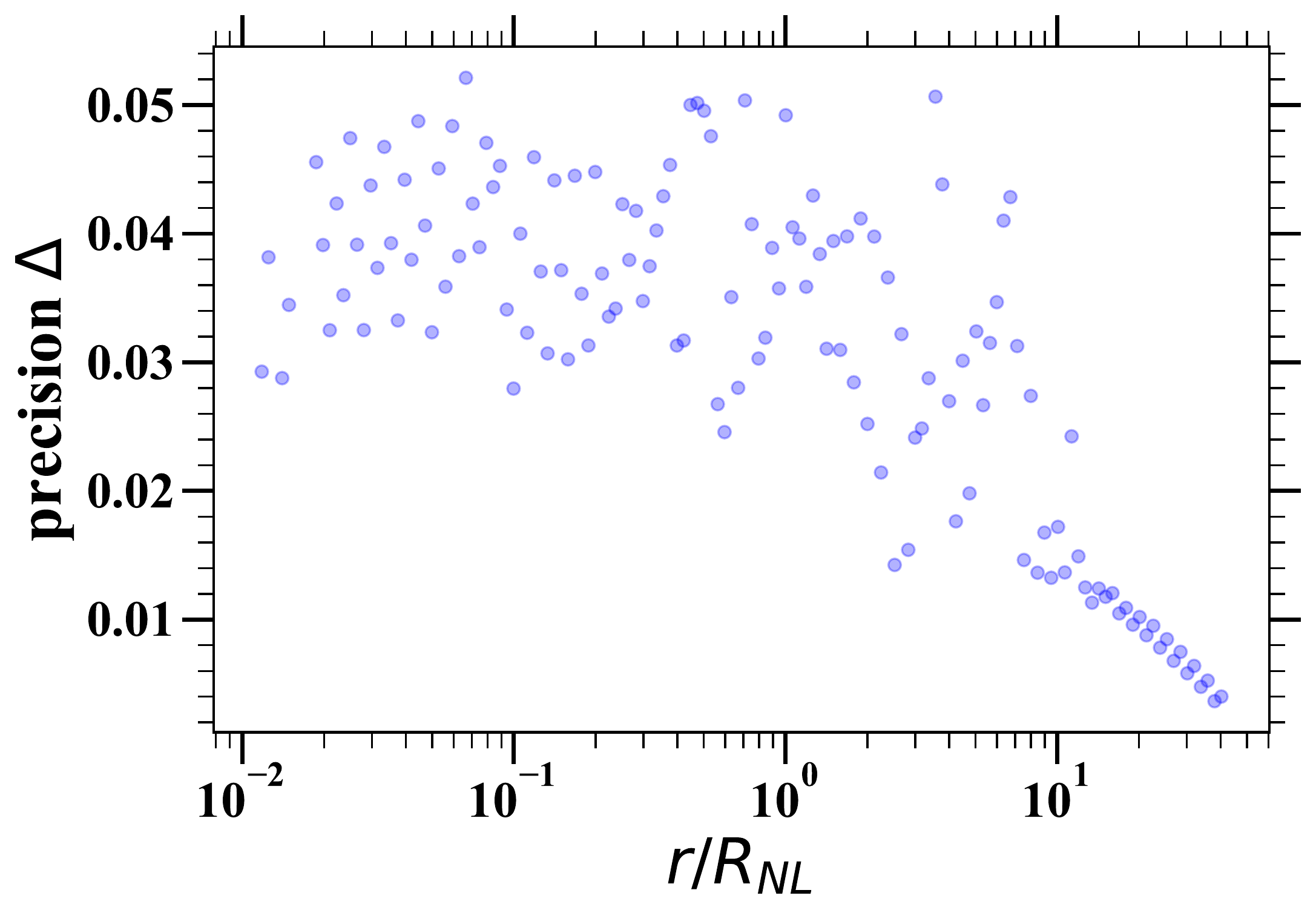}}
\caption{Same quantities as in previous figure, but corresponding to the criterion that the 2PCF in the defined 
regions differ by less than 5 percent from  the converged 2PCF as estimated using the first criterion.
The precision $\Delta$ is calculated in the same way, but relative to the average 2PCF in the new regions.}
\label{resolution-plots-2}
\end{figure*} 

The lower panels in each plot in Figure \ref{fig-convergence} show that the convergence, apparent visually as the flattening of the curves in the upper panels, has a clear signature: in these regions the fractional variation $\Delta\xi/\xi$ 
is not only small but its sign fluctuates about zero. 
This can be used to define converged regions and an estimated converged value for each rescaled bin. The level at which convergence may be obtained is quantified by these fluctuations, which represent a residual level of breaking of self-similarity due to sampling variance of the estimator. To identify the converged regions we 
thus simply identify a minimal number of consecutive snapshots in which $\Delta\xi/\xi$ is below this characteristic level. The vertical dashed lines in the upper panels in Figure \ref{fig-convergence} indicate the regions identified by choosing the threshold level as given by the 
horizontal dashed lines, and requiring at least three consecutive snapshots. The choice of threshold can be conveniently stated as a bound on the
logarithmic derivative $d(\log \xi)/d(\log a)$ since, given that
$\Delta \log_2 a=1/12$, we have $d(\log \xi)/d(\log a) \approx (17.3) \Delta\xi/\xi$. The threshold in Figure \ref{fig-convergence} corresponds 
to  $d(\log \xi)/d(\log a)\approx 0.1$.

The mean 2PCF in each resolved regions gives an estimate for the true 2PCF in the corresponding bin, which we will denote $\xi_{\rm conv} (r/R_{\rm NL})$, We then estimate a ``precision" 
for each bin, denoted $\Delta (r/R_{\rm NL})$, calculated as the maximal fractional difference between any value of the 2PCF in the converged region  and the estimated converged value $\xi_{ \rm conv}(r/R_{\rm NL)}$. We note that
this quantity is thus not an estimated error on the determination of the mean, but an estimated upper bound to the difference between the 2PCF measured in a single snapshot in the resolved regions and the true converged value. The latter corresponds more to a practical definition of precision in the context of $N$-body simulation.   

Finally the estimated converged value $\xi_{\rm conv} (r/R_{\rm NL})$ in each bin
can be used to define temporal regions in which the measured CF is within some 
input fraction $f$ of this value, i.e. we can determine, for each rescaled bin, the 
range of $a/a_0$ for which  
\begin{equation}
1-f < \frac{\xi(r/R_{\rm NL}, a)}{\xi_{\rm conv}(r/R_{\rm NL})} < 1+f\,.
\label{criterion}
\end{equation} \\
Using the points in this range, we can then determine again the precision 
$\Delta(r/R_{\rm NL})$ in the same way as above, taking the new average value 
of the CF in the range as the converged value in each bin.  
When $f$ is significantly larger than the 
initial estimated precision $\Delta (r/R_{\rm NL})$, this new 
estimated precision would be expected to be of order $f$ for all bins.
As we now discuss in detail, we can determine resolved regions for a desired level of
precision, provided it is greater than the limiting intrinsic 
precision inferred from our analysis above.
 
\section{Inferred resolution limits and their interpretation}
\label{Interpretation}

Using the results detailed above, it is straightforward to infer
the comoving sizes of all bins which are resolved at each time, as well as the associated precision.
The results are shown in Figures \ref{resolution-plots-1}  and \ref{resolution-plots-2}. In 
both cases the left panels plots the resolved scales, in units of $\Lambda$, as a function
of time, while the right panel shows the precision $\Delta (r/R_{\rm NL})$. Figure \ref{resolution-plots-1} corresponds
to the tight resolution limits inferred directly from the convergence plots as described in the
previous section. The dashed lines indicate lines of constant $r/R_{NL}$ and constant amplitude 
of the 2PCF, for the indicated values. Figure \ref{resolution-plots-2} corresponds to the less restrictive condition given by using Eq.~\ref{criterion} with $f=0.05$.
The corresponding precision $\Delta (r/R_{\rm NL})$ in the right panel of Figure \ref{resolution-plots-2} is, as would be expected,  more uniform and 
of order $f$, except at the largest linear scales where, as we
will discuss further below, its precision remains high (i.e. $\Delta (r/R_{\rm NL})$ is small) 
at all times for which the 2PCF is measured.

Figures \ref{resolution-plots-1} and \ref{resolution-plots-2} are our central and most important results. 
To our knowledge they provide for the first time in the literature a precise quantification of the resolution of a statistic as a function of time in a cosmological  $N$-body simulation. We now 
discuss in detail the copious information these plots contain, in particular about how the lower cut-off to resolution evolves in time.

Both figures have, to a first approximation, the same principal features.
The evolution of the resolution of the 2PCF is broadly characterized by two regimes.
In the first phase, at early times, the minimal resolution scale is of order $\Lambda$. 
The second phase starts abruptly with a steep decrease of the minimal resolved
scale by almost an order of magnitude (to $\sim \Lambda/5$). 
Intermediate scales converge progressively, somewhat later than the smaller ones, until all scales between a single lower and upper cut-off are resolved. The lower cut-off evolves slowly, decreasing  roughly monotonically. The upper cut-off is 
simply a result of the  mask adopted for our calculation of the 2PCF, except at the very latest times when there are deviations due to finite box size. 

\subsection{Evolution of resolution from $a/a_0 \sim 2$}
The limitation of resolution at early times to scales of order $\Lambda$ 
reflects the imprint of the initial particle grid, which as seen in
the upper panels of Figure \ref{fig-CFs} leads to ``wiggle'' 
features in the 2PCF at early times. The associated 
restriction on the resolved scales differs between Figure \ref{resolution-plots-1} and Figure \ref{resolution-plots-2}
as would be expected given the relative strictness of the 
criteria for resolution used in each case.

The abrupt appearance, at $a \sim 2 a_0$, of resolution at 
smaller scales results from the onset of strongly non-linear gravitational clustering. This efficiently creates mass fluctuations at smaller comoving scales which start to contribute to the correlation signal.
Indeed we observe that this occurs when the resolved 2PCF is of 
order unity, and that $a/a_0 \sim 2$ corresponds to $\sigma_{\rm lin} (\Lambda, a) \sim 1$. The associated scale ($\sim \Lambda/5$) is thus 
simply the typical size of the first significant collapsed structures,
predicted by the simple spherical collapse model to be of this order ($\sim (200)^{1/3}$). We note that given the very significantly smaller value of the force smoothing here ($\epsilon=\Lambda/30$), the clustering at
this scale should not be  influenced at any significant level by 
the force smoothing.

More subtle and unexpected, we see that this process of propagation of 
resolution to smaller scales is not uniform. Instead there is a transition period, extending up to about $\log_2(a/a_0) \approx 1.5$ in Figure \ref{resolution-plots-2}, in which a range of 
intermediate scales are progressively resolved {\it after} scales 
smaller than them are first resolved. The explanation for this behaviour (also seen, albeit less markedly, in the left panel of  Figure \ref{resolution-plots-2}) is simple: it arises
because of the localized nature of the features arising from the 
particle grid 
in the early time 2PCF  
at scales of order $\Lambda$. As noted in  Figure \ref{fig-CFs},
the initial 2PCF has a small bump feature just above $\Lambda$, another at
approximately $2.5\,\Lambda$, as well as a marked dip  at $ 0.5 \,\Lambda$.
In Figures \ref{resolution-plots-1} and  \ref{resolution-plots-2} we see that the intermediate unresolved scales at earlier times are centred 
exactly at these scales. Thus the deviations from the self-similar 2PCF at these scales persist above the chosen threshold level until the contribution to the 2PCF at these scales from the newly forming non-linear structure structures dominates over the initial correlation signal coming from the lattice.  
How this delays the convergence at these specific scales can 
be seen clearly in the plots for the last two bins in Figure \ref{fig-convergence}: these bins are in the range 
where the 2PCF converges just around the time 
when the comoving scale of the bin is of order 
the initial lattice spacing (indicated by the vertical dotted lines).
The slight excess of initial correlation around this scale 
thus blocks the convergence (i.e. degrades the resolution
around this scale until a later time). The same effect is responsible for the 
narrow missing stripe region in the left panel of Figure \ref{resolution-plots-1}. The corresponding bins are located  
between the last two bins shown in Figure \ref{fig-convergence},
where the region picked out as convergent by 
our chosen criteria becomes very sensitive to the precise position and 
amplitude of this bump feature. By using a slightly less constraining 
criterion the resolution plots of Figure \ref{resolution-plots-1}  lose these
features and have the greater regularity of Figure \ref{resolution-plots-2}.
 
\subsection{Precision}

We next comment further on our estimated precision, shown in the right panels of
Figures \ref{resolution-plots-1} and  \ref{resolution-plots-2}. As noted above,
we have defined the precision $\Delta$ as  an estimator for the maximal difference between the measured 2PCF in the resolved regions and the converged value. 

Examining carefully the convergence plots in Figure \ref{fig-convergence}, 
it is straightforward to identify  two distinct regimes in the range
of $r/R_{NL}$ in the right panel of Figure \ref{resolution-plots-1}. 
Starting from the smallest values of $r/R_{NL}$ at which convergence
is obtained, and up to $r/R_{NL} \approx 0.3$, the width of the 
resolved regions grows and the distribution of the values
about the mean appears to be fairly symmetric. The estimated precision, 
below one percent, reflects the points at the extreme of the regions.
Thus it appears, for all but the smallest $r/R_{NL}$ in which there are 
very few points in the resolved regions, to be a conservative upper 
bound  on the precision in most of the region. Starting from values 
of $r/R_{NL}$ between $0.3$ and $0.5$, we see emerge, in addition to 
the effect of the localized features discussed above, a different behaviour:
a small but clear systematic decrease in amplitude as a function of 
time across the resolved regions (as can be seen by examining the sign of $\Delta \xi/\xi$ in the last bins in Figure \ref{fig-convergence} ).
This leads to the observed degraded precision in the right panel of Figure \ref{resolution-plots-1} starting from  about $r/R_{NL} \approx 0.3$. 
The origin of this behaviour is clearly the finite box size: the 
associated comoving scales, in the weakly non-linear regime,
grow with time and become increasingly sensitive to 
missing power at the box scale. The resulting growth of the 
precision continues up to $r/R_{NL} \approx 5$,  when the mask used in our
2PCF calculation starts to limit the selected regions and artificially
depresses the measured precision. 

In summary our conclusion is that, in the resolved regions shown in 
the left panel of Figure \ref{resolution-plots-1}, the {\it precision 
in the non-linear regime is robustly below the percent level}, 
except for the smallest scales  --- conservatively, below 
$r/R_{NL} \sim  0.04$ corresponding to $\xi \sim 250$,   where  
the size of the converged regions is considerably smaller and the 
estimate of precision less reliable. At larger scales --- in the very
weakly non-linear and linear regime --- precision is, on the other 
hand degraded by finite box size effects and possibly as large
as two percent in the corresponding regions. 

The precision corresponding to the resolved regions in 
Figures \ref{resolution-plots-2} is simpler to understand: these
regions are constructed by definition to have a target precision
of five percent. The strong suppression below this value 
at large scales is, as in  the right panel of Figures \ref{resolution-plots-1},
due simply to the upper cut-off imposed by the mask in our
2PCF calculation. The significant dispersion around 
$\Delta \sim 0.04$ in the rest of the plot arises because 
the precision has been defined with respect to the average
2PCF in the new regions. 
 
\begin{figure*}
\centering\resizebox{8.5cm}{!}{\includegraphics[]{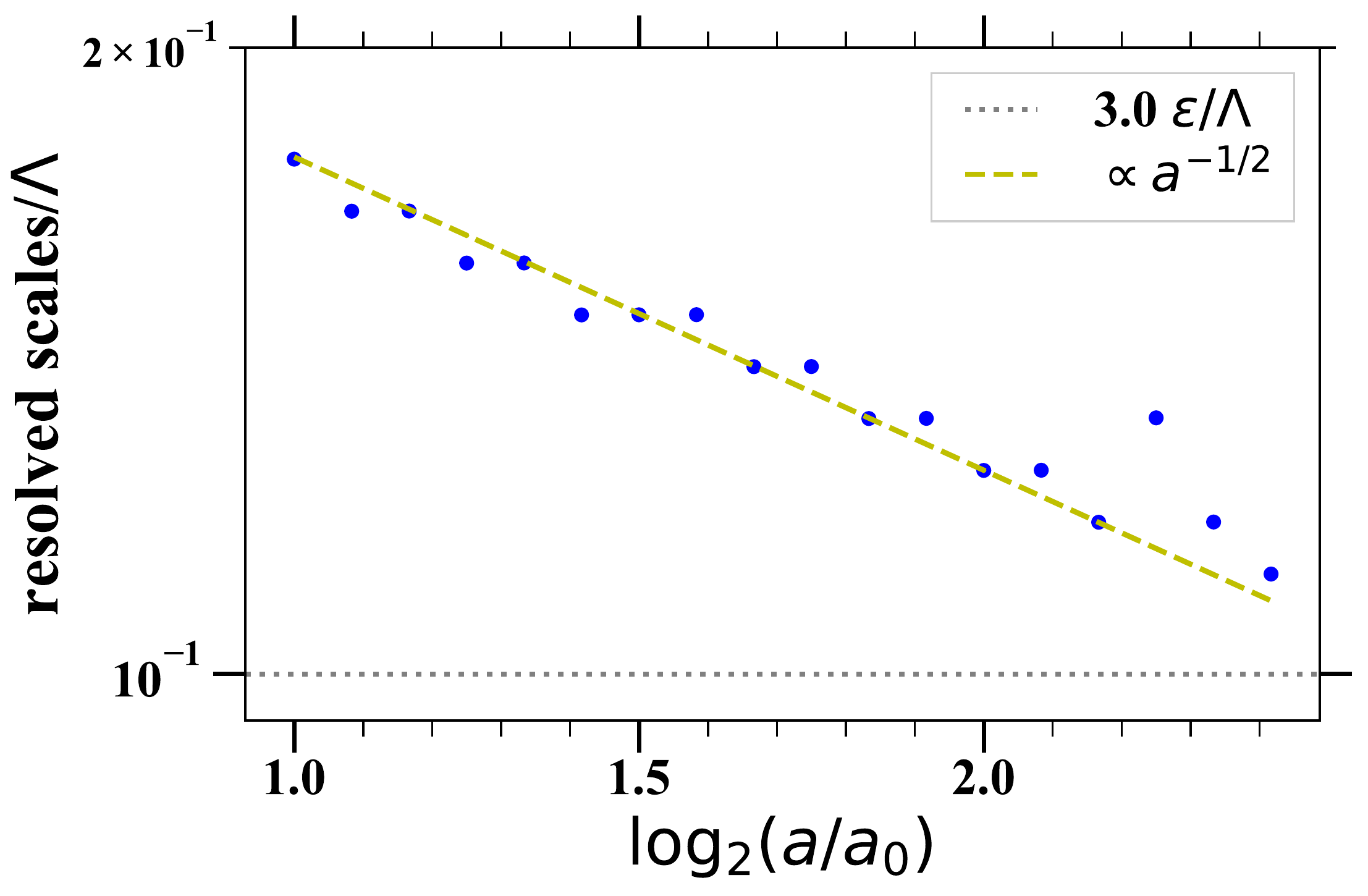}}
\centering\resizebox{8cm}{!}{\includegraphics[]{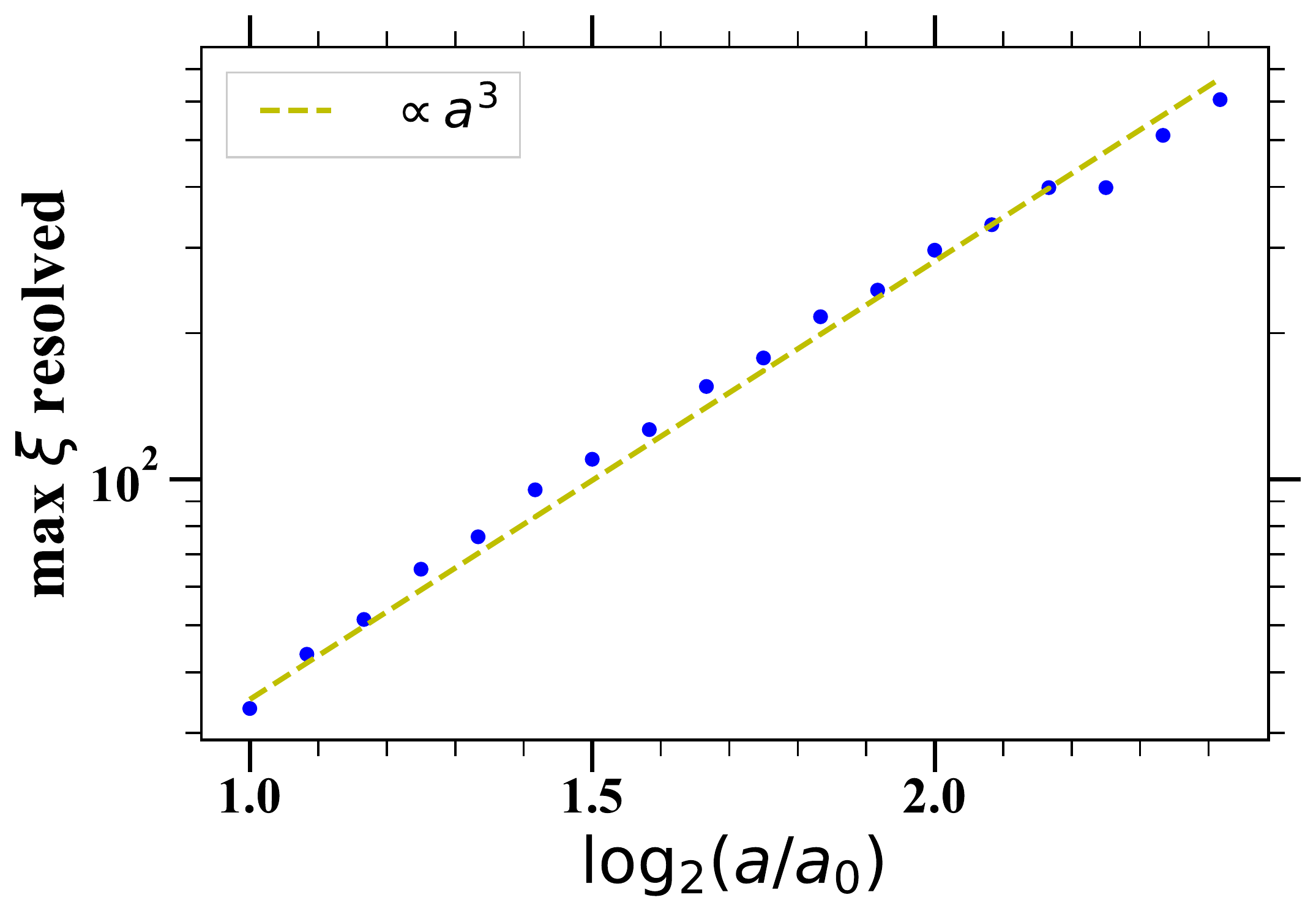}} 
\caption{Left panel: Minimum resolved scale as a function of time, at later times, as determined using the same criteria as in Figure \ref{resolution-plots-2}; the scale $3 \epsilon/\Lambda$ is shown for reference, and the dashed straight line (normalized by eye) decays as $a^{-1/2}$ indicative of two-body relaxation. Right panel:  The corresponding maximum resolved 2PCF as a function of time; the dashed straight line is proportional to $a^{3}$ (normalized by eye), the behaviour corresponding to fixed density in physical coordinates at the minimal resolved scale.}
\label{resolution-plots-small-scales} 
\end{figure*}


\subsection{Evolution of the lower cut-off at late times}

We now consider more closely the evolution of the lower cut-off
to the resolved scales at later times.  The physical effects potentially contributing to this evolution  
are numerous (e.g. two-body interactions, finite $N$ instabilities, unresolved merging) and, in such a 
complex out-of-equilibrium system, they are poorly understood. Nevertheless in isolated $N$-body 
systems in dynamical equilibrium, there is excellent 
numerical evidence that the simple description of finite $N$ relaxation as due essentially to incoherent two-body 
scatterings originally proposed by Chandrasekhar is accurate
(\cite{Farouki+Salpeter_1982,Farouki+Salpeter_1994, Theis_1998, Marcos_etal_2017}). In the context of cosmological $N$-body 
simulations, several studies (e.g. \cite{power2003inner,ludlow_etal_2019}) have advocated that 
these kind of effects appear to model well resolution-dependent behaviours in the inner parts of halos (see also \cite{knebe_etal_2000}, which  focusses on issues associated with poor numerical integration of
strong two-body deflections). Given that the strongly non-linear 
2PCF is dominated by halos, we might expect this to be the 
relevant mechanism here. 

The left panel of Figure \ref{resolution-plots-small-scales} shows
the smallest resolved scale as a function of time starting 
from times at which non-linear scales ($\xi > 10$) are resolved, for
the same criteria as used in the first panel in Figure \ref{resolution-plots-2}.
The dashed line decreasing as $a^{-1/2}$ provide a good fit to the average trend.
The right panel in Figure \ref {resolution-plots-small-scales}  shows the corresponding 
maximal resolved value of the 2PCF, for exactly the same data points. In this case
a line proportional $a^{3}$ is shown and clearly provides a good phenomenological
fit to the data. As we now explain, these functional behaviours 
appear to be very consistent with the conclusion that this evolution is indeed 
predominantly set by two-body collisionality.

A comoving resolution scale decreasing in proportion to $a^{-1/2}$ corresponds to a resolution scale 
{\it in physical coordinates}, $r_{\rm res}$, growing in proportion to  $a^{1/2}$. 
Given that $a \propto t^{2/3}$, the observed behaviour corresponds to a scaling
 $r_{\rm res}^3 \propto t$.  Let us assume, as in \cite{power2003inner, ludlow_etal_2019}, 
that two-body collisions inside a halo leads to progressive relaxation as a function of radius 
$r$ on an effective time-scale 
\begin{equation}
t_{\rm 2-body} (r) \sim \frac{N(<r)}{\sqrt{G \rho_{\rm av} (<r)}}
\label{2body_time}
\end{equation}
where $N(<r)$ is the number of particles enclosed, and 
$\rho_{\rm av} (<r)$ is the average physical density in the region.
(For simplicity, we have neglected the logarithmic correction to
this behaviour in $N$ or the softening $\epsilon$).
Identifying now $t=t_{\rm 2-body} (r_{\rm res})$, we obtain
\begin{equation}
r_{\rm res}^3  \sim \frac{N(<r_{\rm res})}{\sqrt{G \rho_{\rm av} (<r_{\rm res})}}\,.
\label{2body-condition}
\end{equation}
As $N(<r_{\rm res})=(4\pi/3) \rho_{\rm av} (<r_{\rm res}) r_{\rm res}^3$,
Eq. (\ref{2body-condition}) is consistent on condition 
that $\rho_{\rm av} (<r_{res})$  is constant {\it in physical coordinates}.
As seen in the right panel of Figure \ref{resolution-plots-small-scales}, the 2PCF at the resolution 
scale grows approximately in proportion to $a^3$, which, as the comoving density also
grows in proportion to $a^3$, is exactly the required behaviour.
In other words, the lower limit to resolution appears to evolve at scales
where the local physical density is constant. This also further justifies
the use of the simple relation Eq. (\ref{2body_time}), which is derived
assuming stationarity of the mass distribution.
The validity of such an approximation in the context of cosmological
clustering corresponds to so-called stable clustering. This apparent
association between the evolution of resolution at small scales
and stable clustering has been previously noted in \cite{benhaiem_etal_2017}.
A detailed analysis of the evidence for the validity of the stable clustering
approximation at small scales in our simulations will be reported in a 
separate forthcoming work.

Our conclusion here is based on several
assumptions and simple approximations. It may be that processes such
as unphysical merging like that analysed in detail 
by \cite{vanDenBosch_et_al_2018} and  \cite{vanDenBosch+ogiya_2018}, 
and which can increase coupling to unresolved fluctuations at smaller 
scales, could be partly or wholly responsible for these behaviours.
Further detailed 
analysis --- in particular of any dependence of this evolution
on force softening --- will be undertaken elsewhere in search of a more
definitive conclusion.

\section{Inferences for simulations of non-scale free cosmologies}
\label{LCDM}

\begin{figure}
\centering\resizebox{8cm}{!}{\includegraphics[]{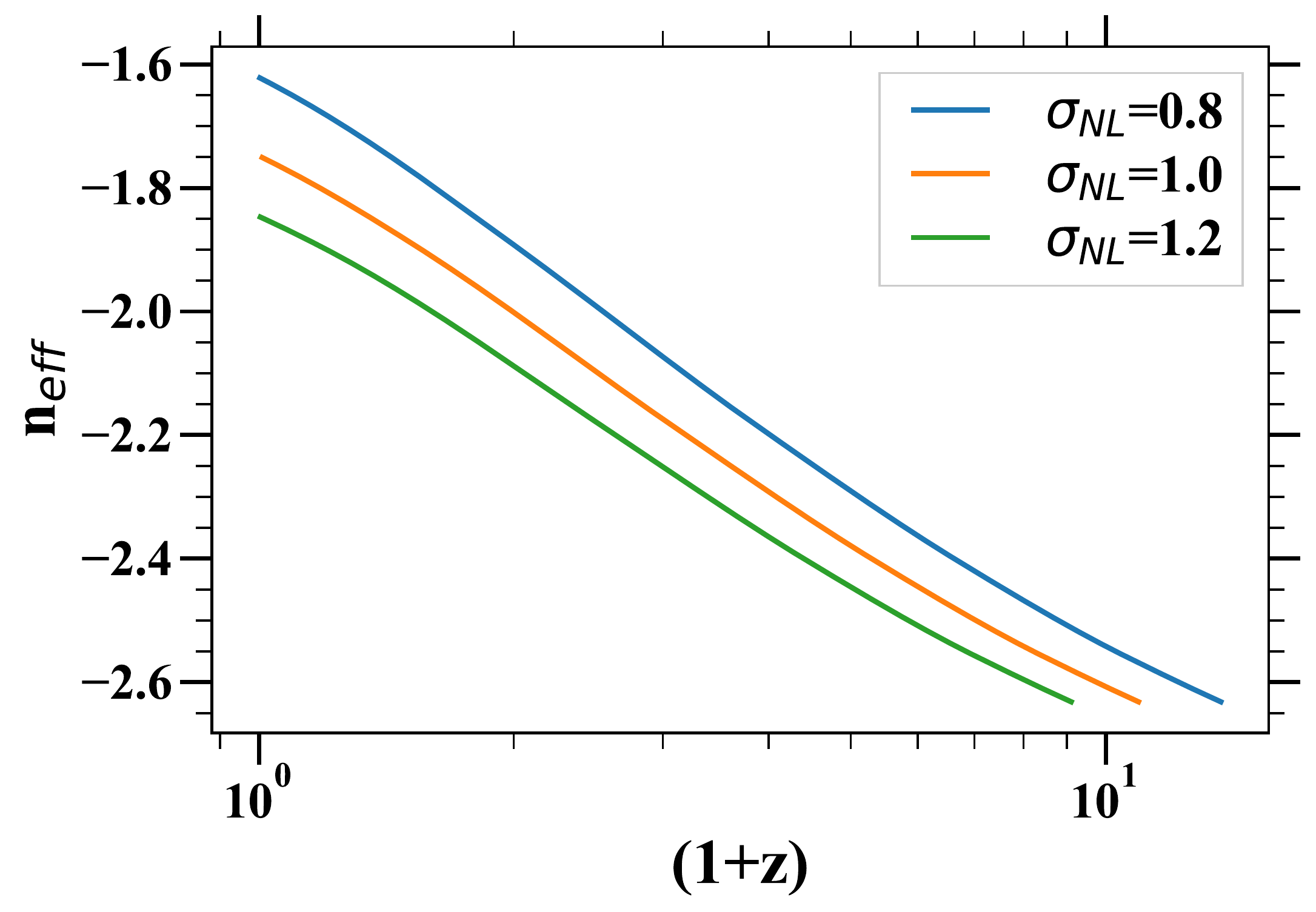}}
\caption{Effective $n$ for the LCDM (``Planck 2013'') model inferred from the logarithmic slope of its linear top-hat variance, at a scale $R$ defined by $\sigma_{lin}(R,z)=\sigma_{NL}$. }
\label{neff-LCDM}
\end{figure}

\begin{figure}
\centering\resizebox{8cm}{!}{\includegraphics[]{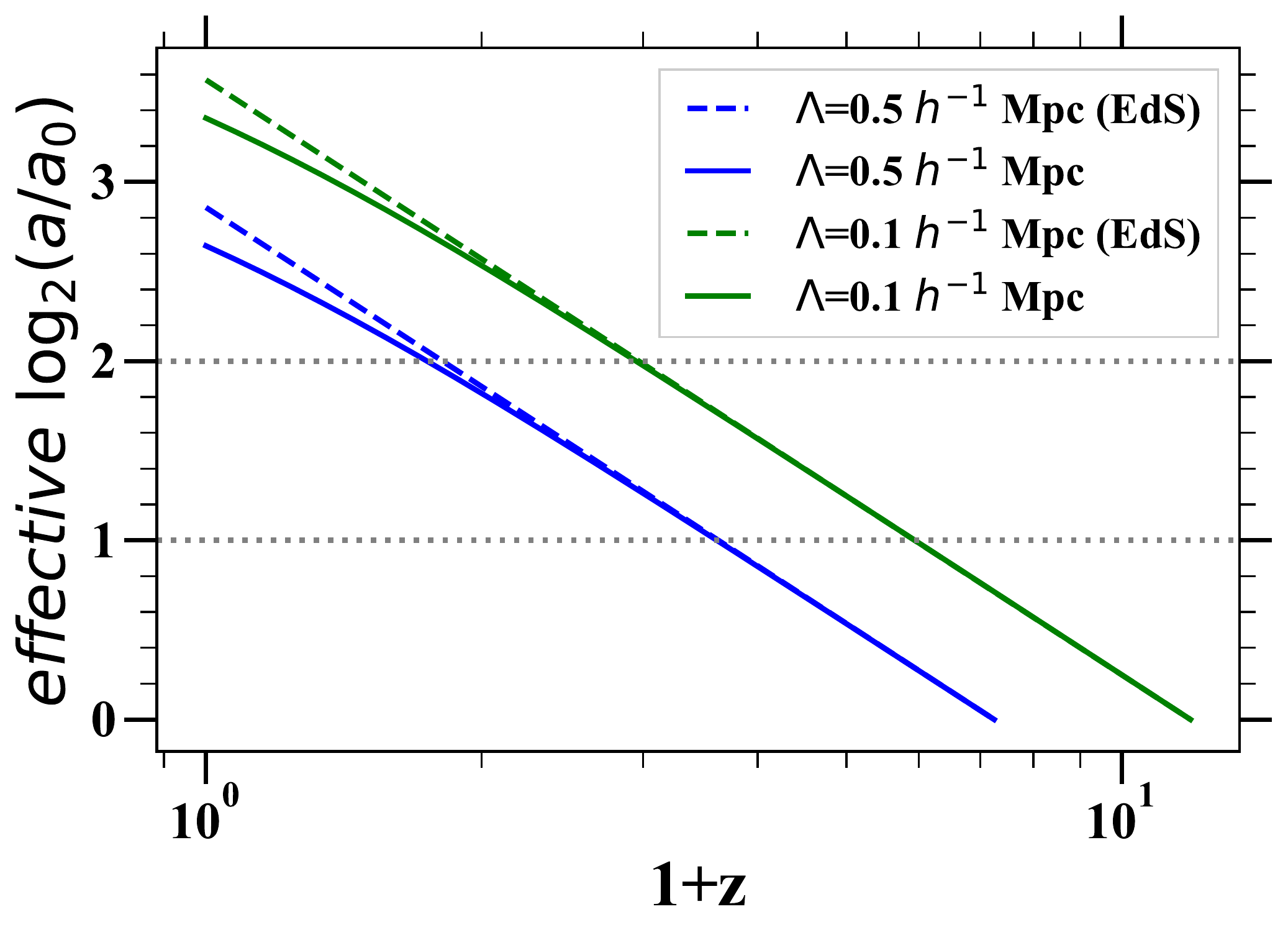}}
\caption{Solid lines: Values of $\log_2(a/a_0)$ to be used in Figures \ref{resolution-plots-1} and \ref{resolution-plots-2} to infer
lower limits to resolution in a LCDM simulation as a function of red-shift, for different physical values of the initial grid spacing, using the approximation described in the text. Dashed lines: values obtained for same quantity in the approximation 
of the LCDM expansion as Einstein-de-Sitter. }
\label{LCDM-mapping}
\end{figure}

Our underlying motivation is to quantify the resolution of simulations not of scale-free models, but of non-scale free cosmologies such as LCDM or variants of it.  We have intentionally framed our characterization of resolution --- given in the form of the left panels of Figures \ref{resolution-plots-1} and \ref{resolution-plots-2} --- so that it can potentially be ``mapped" easily onto such simulations. The resolved comoving scales are given, in units of $\Lambda$, and as a function of $(a/a_0)$ only. Given the input parameters of an LCDM simulation, $a_0$, as defined in  Eq. (\ref{a0-ref}), can be calculated unambiguously given the model parameters and $\Lambda$, and thus one can directly infer an estimate resolution at a function of red-shift in a simulation.
The limitation on the reliability of this procedure comes from the facts that (1) the power spectrum of an LCDM cosmology is not a simple power law, and (2) the expansion history deviates from the Einstein-de-Sitter model at late times.

For the former, the question is thus how much the resolution depends on the initial spectrum: 
while $a_0$ does not depend on it by construction, any of the characteristics of the measured resolution may do
so, even strongly. Such a dependence can be expected to be quantifiable in principle by studying scale-free 
models with exponents $n$, in range which can be inferred from Figure \ref{neff-LCDM},
which shows an effective exponent $n_{eff}$ characterizing fluctuations in a typical LCDM model 
(``Planck 2013'', see \cite{planck_2013}) as a function of red-shift, as inferred from the logarithmic slope of 
$\sigma_{lin}$ at a scale $R(z)$ defined by $\sigma_{lin} (R,z)=\sigma_{NL}$, for a few 
different values of $\sigma_{NL}$ are shown\footnote{The results in this figure and the next one have been obtained using the Python toolkit {\it Colossus} \citep{colossus_2018}.}. We will report such a study in forthcoming work.
However it is clear that, while resolution effects associated 
with finite size box size are expected to depend strongly on the 
effective exponent,  we do not expect a very significant dependence of the
resolution at small scales on the initial spectrum. Indeed the factors which are relevant in fixing the relevant scales and behaviours are in principle very insensitive to the model. As discussed above, they are fixed essentially by $\Lambda$, and
by very general considerations about the processes of gravitational collapse and
two-body relaxation. For the latter, our simple analytical model gives a 
behaviour which is model independent.

For what concerns the resolution at small scales and the deviation from Einstein-de-Sitter of the LCDM cosmology,
we note that this reduces --- assuming our attribution of its evolution predominantly 
to two-body collisionality is correct --- to the question of how this evolution depends
on the model. If the inferred late time evolution is expressed as a function of time,
i.e. as $r_{res}  \propto t^{-1/3}$, it is then model independent, and, neglecting 
model dependence of any other feature, our result for $n=-2$ can be mapped onto 
LCDM without ambiguity. Using this method Figure \ref{LCDM-mapping} shows in 
a convenient form how the resolution, as a function of red-shift, in an LCDM simulation
with a specified initial grid spacing and given model (here again assumed to be 
Planck 2013), can then be inferred approximately from the results given in
Figures \ref{resolution-plots-1} and \ref{resolution-plots-2}: it gives, as a 
function of red-shift, the value of $\log_2 (a/a_0)$ at which the resolution
should be inferred in these figures. For each given $\Lambda$ we
show both the inference with and without the  small modification 
taking account of the non-Einstein-de-Sitter cosmology. To infer resolution at low
red-shift in a very large cosmological simulation supposes evidently
a simple extrapolation of the late time evolution we have found.
As we have done here the LCDM simulation must 
be tested  separately for its convergence with respect to numerical
integration parameters.

\section{Discussion} 

Our analysis has shown that the self-similarity of a scale-free simulation is an excellent tool to quantify the resolution of cosmological $N$-body simulations. Scale-free simulations have been performed by numerous groups in the past, but with the exception of a few recent works \citep{orban2013keeping, benhaiem2013self, benhaiem_etal_2017}, self-similarity tests have been applied only quite qualitatively (e.g. comparing plots of 
physical quantities at a few different low red-shifts). We have used it here to give, for the first time, a precise quantitative 
characterization of the evolution of resolution of a statistic measured 
in an $N$-body simulation. In doing so we have tested the accuracy of the
\Abacus code, showing that it can measure the self-similar 2PCF to an 
accuracy well below the percent 
level in a range of scale. Requiring precision at the level of a few percent, we have been able to follow the evolution
of the associated resolution scale over a range of scale-factor sufficiently large to allow us even to determine approximately
their functional behaviours. The latter appear to be very consistent with the hypothesis that resolution at later times is
limited by two-body collisionality inside halos that are approximately stable in physical coordinates, but we have not 
excluded definitively that other processes may be involved or even dominate.

The question of the small-scale resolution of the 2PCF, and the power spectrum (PS), measured in $N$-body simulations has been subject of discussion and even controversy in the literature. Given the satisfaction of appropriate criteria for parameters controlling the accuracy of the $N$-body integration, it is evident that resolution must be controlled by the parameters $\Lambda$ and $\epsilon$. An  
often used nomenclature which refers to $\epsilon$ as {\it the spatial resolution} of an $N$-body simulation (while $\Lambda$ is referred to as the {\it mass resolution}) corresponds to an apparently common idea that it is $\epsilon$ which determines the lower cut-off to spatial resolution.  On the other hand, going back to work
by \cite{melott_1997,splinter_1998}, even the claim that resolution can be extended
below the scale $\Lambda$ at all has been discussed and studied by several authors
who reach quite different conclusions \citep{ knebe_etal_2000, romeo_etal_2008, discreteness3_mjbm}.

The methodology we have developed here is, we believe,  one which can potentially resolve definitively these issues. The specific results we have presented for the 2PCF in the $n=-2$ model show how the spatial resolution at small scales is fundamentally determined  by the scale $\Lambda$. The initial scale of resolution is initially fixed by $\Lambda$, and then evolves very non-trivially in a way which depends on both the dynamics of gravitational collapse itself and on processes like two-body relaxation. The gravitational softening $\epsilon/\Lambda=1/30$ used here is sufficiently small that it is at all times, except perhaps in the very last few snapshots, very much smaller that the lower cut-off to resolution we have determined. If a significantly larger value of $\epsilon/\Lambda$ is used, we can anticipate that it will set an 
absolute lower limit on resolution starting from a correspondingly earlier time. Conversely using a smaller $\epsilon/\Lambda$ than that we have employed would not be expected to modify the resolution (although the numerical cost of accurate integration 
of trajectories may become prohibitive).
In a forthcoming
article we will test these conclusions more directly by studying simulations of the same scale free model with a range of different values of the force softening. 

The results we have presented here are an illustration 
of the method we have developed, showing how powerful a tool it provides to understand and quantify resolution limits on simulations. Here we have used only a single scale-free model, simulated at fixed values of the discretization parameters, 
and a single statistic, but the method can be applied much more broadly. Doing so will allow us also to compare our results more directly and quantitatively to those of numerous other works in the literature which have used different methods to constrain the precision of  cosmological $N$-body simulations: in particular such studies have focussed
mostly on the precision  of the power spectrum measured in cosmological $N$-body simulations (see e.g. \cite{heitmann_et_al_2014, smith_et_al_2014, klypin+prada_2019,euclid_collab_2019a, smith+angulo_2019, cataneo_2019}) and of halo statistics (\cite{power2016spurious, klypin_et_al_2015, ludlow_etal_2019, vanDenBosch_et_al_2018, vanDenBosch+ogiya_2018, green+van_den_bosch_2019}).
We will extend our study to both kinds of statistics and explore how the method allows us, for example, to determine quantitatively how choices about initial conditions and force softening affect resolution and how these choices may be optimized.      

\section*{Acknowledgements} 

M.J. warmly thanks the Institute for Theory and Computation for hosting him as a sabbatical visitor for the academic year 2017-18 during which this collaboration was initiated, and acknowledges David Benhaiem, Bruno Marcos and Francesco Sylos Labini for collaboration and many useful discussions on related subjects. We thank Benedikt Diemer for providing us with data from his scale-free simulations which have very useful for purposes of comparison, and Mika\"{e}l Leroy for useful discussions. 
D.J.E. is supported by the U.S. Department of Energy grant DE-SC0013178 and as a Simons Foundation Investigator. L.H.G. is also supported by the Simons Foundation.

\section*{Data availability} 

The data reported in the article are provided in csv format in the online supplementary material.


\end{document}